\documentclass[fleqn,usenatbib]{mnras}

\usepackage[T1]{fontenc}

\DeclareRobustCommand{\VAN}[3]{#2}
\let\VANthebibliography\thebibliography
\def\thebibliography{\DeclareRobustCommand{\VAN}[3]{##3}\VANthebibliography}

\usepackage{graphicx}	
\usepackage{amsmath}	
\usepackage{amssymb}	
\usepackage{ulem}

\usepackage{newtxtext,newtxmath}

\newcommand{\lya}{Ly$\alpha$ }

\newcommand{\appropto}{\mathrel{\vcenter{
  \offinterlineskip\halign{\hfil$##$\cr
    \propto\cr\noalign{\kern2pt}\sim\cr\noalign{\kern-2pt}}}}}

\title[Memory of reionization in Ly$\alpha$ cross 21 cm]{Reionization relics in the cross-correlation between the Ly$\alpha$ forest and 21 cm intensity mapping in the post-reionization era}

\author[Montero-Camacho et al.]{
Paulo Montero-Camacho,$^{1,2}$\thanks{E-mail: pmontero@pcl.ac.cn (PMC)} Catalina Morales-Gutiérrez,$^{3,4}$ 
Yao Zhang,$^{2}$
Heyang Long,$^{5,6}$ and
Yi Mao$^{2}$
\\
$^{1}$Department of Mathematics and Theory, Peng Cheng Laboratory, Shenzhen, Guangdong 518066, China\\
$^2$Department of Astronomy, Tsinghua University, Beijing 100084, China\\
$^3$Department of Physics, University of Costa Rica, 11501 San José, Costa Rica.\\
$^4$Space Research Center (CINESPA), University of Costa Rica, 11501 San José, Costa Rica.\\
$^5$Department of Physics, The Ohio State University, 191 West Woodruff Avenue, Columbus, OH 43210, USA\\
$^6$Center for Cosmology and AstroParticle Physics (CCAPP), The Ohio State University, 191 West Woodruff Avenue, Columbus, OH 43210, USA
}

\date{Accepted XXX. Received YYY; in original form ZZZ}

\pubyear{2024}

\begin{document}
\label{firstpage}
\pagerange{\pageref{firstpage}--\pageref{lastpage}}
\maketitle

\begin{abstract}
The tumultuous effects of ultraviolet photons that source cosmic reionization, the subsequent compression and shock-heating of low-density regions, and the modulation of baryons in shallow potential wells induced by the passage of ionization fronts, collectively introduce perturbations to the evolution of the intergalactic medium in the post-reionization era. These enduring fluctuations persist deep into the post-reionization era, casting a challenge upon precision cosmology endeavors targeting tracers in this cosmic era. Simultaneously, these relics from reionization also present a unique opportunity to glean insights into the astrophysics that govern the epoch of reionization. In this work, we propose a first study of the cross-correlation of \lya forest and 21 cm intensity mapping, accounting for the repercussions of inhomogeneous reionization in the post-reionization era. We investigate the ability of SKA $\times$ DESI-like, SKA $\times$ MUST-like, and PUMA $\times$ MUST-like instrumental setups to achieve a high signal-to-noise ratio (SNR) in the redshift range $3.5 \leq z \leq 4$. Moreover, we assess how alterations in integration time, survey area, and reionization scenarios impact the SNR. Furthermore, we forecast the cross-correlation's potential to constrain cosmological parameters under varying assumptions: considering or disregarding reionization relics, marginalizing over reionization astrophysics, and assuming perfect knowledge of reionization. Notably, our findings underscore the remarkable capability of a futuristic PUMA $\times$ MUST-like setup, with a modest 100-hour integration time over a 100 sq. deg. survey, to constrain the ionization efficiency error to $\sigma_\zeta = 3.42
$. 
\end{abstract}




\section{Introduction}
Cosmic reionization, the transformative phase during which our Universe shifted from predominantly neutral to highly ionized, is estimated to have occurred approximately around redshift $z \sim 8$ \citep[e.g. ][]{2020A&A...641A...6P}. While the overarching mechanisms driving cosmic reionization are broadly understood \citep{2016ASSL..423.....M,2019cosm.book.....M}, the lack of direct observations introduces significant uncertainties, particularly regarding the timeline of reionization \citep{2020MNRAS.491.1736K,2022ApJ...937...17G,2023ApJ...942...59J,2023arXiv231106348R}. With the advent of the James Webb Space Telescope \citep[JWST;][]{2006SSRv..123..485G} our knowledge of the sources of ultraviolet photons responsible for reionization is likely to increase dramatically. Moreover, reionization unfolds as a markedly inhomogeneous inside-out process \citep{2008ApJ...675....8L,2009MNRAS.394..960C} wherein denser regions undergo ionization first since sources of ultraviolet photons are expected to predominantly be situated within these dense environments.

The post-reionization era is a treasure trove of cosmological information. The relatively high redshifts characteristic of this period not only facilitate the exploration of dark matter candidates \citep{2020JCAP...04..038P,2022MNRAS.tmp.3519P} but also offer a unique opportunity to study cosmic evolution and structure formation \citep{2009MNRAS.397.1926W,2009JCAP...10..030V} before the onset of nonlinearities imposes significant constraints on the observational landscape.

The Lyman-$\alpha$ (Ly$\alpha$) forest, the absorption features observed in the spectra of background quasars, stands as a pivotal tool for exploring the Universe in the post-reionization era. Its applications extend to the study of \ion{H}{I} and \ion{He}{II} reionization \citep{2009ApJ...706L.164C,2020MNRAS.496.4372U,2020MNRAS.499.1640M}, the investigation of the matter power spectrum on scales beyond the reach of galaxy surveys \citep{2003AIPC..666..157W}, constrain the evolution of the Universe \citep{2020ApJ...901..153D}, and the inference of cosmological parameters \citep{2019JCAP...07..017C}. Moreover, the forest also offers a unique window into the impact of neutrino masses \citep{2017JCAP...06..047Y}. However, obtaining reliable cosmological measurements of the \lya forest at high redshifts proves challenging due to the sparse sampling of quasars \citep{2020RNAAS...4..179Y,2021MNRAS.508.1262M,2023ApJ...944..107C}, particularly regarding potential measurements of the 3D flux power spectrum\footnote{Note that with the current generation of available instruments, a significant increase in the number of high-redshift quasars is occurring \citep{2023arXiv230201777Y}; however, despite these developments, achieving the statistical density required to reduce the mean separation between line-of-sights for a broadband 3D power spectrum analysis remains a formidable task that may be achieved with the current generation of available instruments -- see \cite{2023arXiv231009116K} for recent efforts in this direction}.  

Intensity mapping (IM) involves a trade-off of sacrificing angular resolution to concentrate on the integrated emission from unresolved sources. Analogous to the \lya forest, IM utilizing the 21 cm hyperfine transition of hydrogen presents a versatile probe capable of delving into various aspects of cosmology in the post-reionization era. It can probe \ion{H}{I} reionization \citep{2023MNRAS.525.6036L}, explore early universe cosmological parameters (e.g. primordial non-gaussianity), and constrain the evolution of the universe and structure formation \citep{2019JCAP...06..025C,2015ApJ...803...21B,2009MNRAS.397.1926W,2009JCAP...10..030V}. Additionally, it can investigate the nature of dark matter \citep{2015JCAP...07..047C} and more \citep{2020PASA...37....7S, 2018arXiv181009572C}. 

Nevertheless, the full potential of this observable is impeded by foregrounds, such as galactic synchrotron emission, which surpass the amplitude of the cosmological signal by several orders of magnitude \citep{2015aska.confE..35W}. Despite dedicated efforts to mitigate foreground effects \citep[see e.g. ][]{2023ApJ...945...38Z,2024arXiv240711296D}, a conservative perspective suggests that confirming the cosmic origin of the signal might necessitate cross-correlation with another probe\footnote{Notably, the CHIME collaboration recently detected cosmological 21 cm emission at $z \approx 1$ through cross-correlation with eBOSS tracers \citep{2023ApJ...947...16A}.}. This cross-correlation typically considers high-$z$ galaxies \citep{2007ApJ...660.1030F,2023ApJ...944...59L,2023MNRAS.525.1664H} as the additional tracer.

The intense heating of the intergalactic medium (IGM) during cosmic reionization triggers a substantial increase in the IGM temperature, reaching a few times $10^4$ K. As ionization fronts propagate, inducing a rise in temperature, low-density regions experience shocks that both heat and compress the gas. These shocks, which originate in denser regions due to the increase in Jeans mass with temperature, will push gas in minivoids to higher adiabats compared to gas in denser environments \citep{2018MNRAS.474.2173H} -- such as minihalos. This high-entropy mean-density gas is then responsible for the long-lasting impact of reionization in the \lya forest \citep{2019MNRAS.487.1047M, 2020MNRAS.499.1640M}, essentially constituting the long-lasting \textit{memory of reionization} at $z \sim 2$ in the forest. Simultaneously, reionization exerts influence over the baryon abundance within a given halo \citep{2022MNRAS.513..117L}. The thermal kick resulting from the passage of an ionization front expels some baryons from the halo, particularly for halos with shallow potential wells. This modulation is responsible for long-lasting reionization relics in 21 cm IM, persisting up to $z \sim 3$ \citep{2023MNRAS.525.6036L}. 

Therefore, both the \lya forest and \ion{H}{I} 21 cm intensity mapping exhibit broad-spectrum contamination originating from the imprints of cosmic reionization. This contamination is especially pronounced at large scales and high redshifts in the post-reionization era, where the lingering effects of reionization are more pronounced due to less time to dissipate the additional injected energy \citep{2020MNRAS.499.1640M}. Managing this systematic is not merely a challenge essential for obtaining cosmological information free from bias in the post-reionization era, but it also represents a novel opportunity to gain insights into the intricate processes governing the reionization history of the Universe \citep{2021MNRAS.508.1262M}.

The observational programs of these two probes are at different developmental stages. The \lya forest has achieved commendable SNR, e.g.  eBOSS \citep{2019JCAP...06..025C,2020ApJ...901..153D} and DESI early \citep{2023arXiv230606312R,2023JCAP...11..045G,2023MNRAS.526.5118R} and year-1 results \citep{2024arXiv240403001D}. In stark contrast, the 2-point functions of the cosmological 21 cm signal at higher redshifts ($z > 2$) are currently constrained only by upper limits \citep[see e.g., ][]{2023arXiv231105364M,2022ApJ...925..221A,2020MNRAS.493.1662M}. Despite this, because of the inherent challenges in measuring the 21 cm auto-power spectrum, there is considerable interest in the cross-correlation of these two distinct tracers as a promising avenue for cosmological studies. 

The \lya forest $\times$ \ion{H}{I} 21 cm IM cross-correlation was initially proposed in \cite{2011MNRAS.410.1130G} as a robust and independent probe of the post-reionization IGM. \cite{2017JCAP...04..001C} underscored the reduced sensitivity of this cross-correlation to foreground contamination, emphasizing its potential to untangle degeneracies inherent in modeling parameters. Furthermore, the \lya $\times$ 21 cm cross-correlation exhibits promising prospects for constraining dark energy \citep{2021JCAP...02..016D} and $f(r)$ gravity models, particularly when coupled with cross-correlations involving cosmic microwave background (CMB) lensing \citep{2023JApA...44....5D}. Moreover, this observable has also emerged as a potential avenue for constraining the nature of dark matter \citep{2019JCAP...12..058S}. 

Regarding the observability of this cosmological probe, the Owens Valley Widefield Array (OWFA) \citep{2014JApA...35..157A,2015JApA...36..385B} and a spectroscopic instrument such as BOSS/eBOSS were anticipated to achieve a robust 6$\sigma$ detection at $z=3.35$ with 200 hours of integration \citep{2018JCAP...05..051S}. Meanwhile, the Square Kilometre Array Phaew 1 Mid-frequency (SKA1-Mid) in conjunction with an eBOSS-like survey could reach a peak SNR of 15 at $z = 2.5$ \citep{2015JCAP...08..001G}. 

However, the existing studies/forecasts have yet to incorporate the enduring impact of inhomogeneous reionization on the post-reionization IGM. In this work, we delve into the repercussions of these reionization remnants in the \lya forest $\times$ \ion{H}{I} 21 cm IM cross-correlation within the redshift range $3.5 \leq z \leq 4$. We specifically target this range because of the heightened abundance of \lya spectra compared to higher redshifts and the pronounced strength of reionization relics in the forest during this period \citep{2020MNRAS.499.1640M}. Nevertheless, it is crucial to note that the impact of reionization imprints will remain significant for this cross-correlation in the broader redshift range of $ 3\lesssim z < 6$. 

The rest of this work is organized as follows. In section \ref{sec:model}, we outline our model for capturing the impact of reionization in the post-reionization IGM. We present the simulations required to compute the reionization relics in section \ref{sec:sims}. The instrumental configurations considered in this study are introduced in section \ref{sec:tele}. Section \ref{sec:results} presents our findings regarding the effectiveness of the instrumental setups in detecting the \lya $\times$ 21 cm cross-correlation. Additionally, we consider deviations in survey design and strategy, providing insights into possible gains. In section \ref{sec:fish}, we demonstrate the importance of accounting for this novel effect through three distinct Fisher forecasts. Finally, we summarized our findings in \ref{sec:sum}. 

Throughout this work, we use $h = 0.6774$, $\Omega_b h^2 = 0.0223$, $\Omega_c h^2 = 0.1188$, $A_s = 2.148 \times 10^{-9}$, and $n_s = 0.9667$. In agreement with the `\textit{TT + TE + EE + lowP + lensing + ext}' cosmology from \cite{2016A&A...594A..13P}. Furthermore, we use the values of $b_{\ion{H}{I}}$ and $\Omega_{\ion{H}{I}}$ reported in \cite{2015MNRAS.452..217C}. In contrast, we obtain the bias and RSD parameters for the forest from \cite{2015JCAP...12..017A}.

\section{Modeling the cross-correlation of \lya and 21 cm IM with reionization relics}
\label{sec:model}
In this work, we consider the inclusion -- at first order -- of the memory of reionization, that is the long-lasting impact of inhomogeneous reionization, in both \lya flux fluctuations and \ion{H}{I} 21 cm fluctuations. We examine the repercussions of this inclusion when embedded into the \textit{traditional} cross-correlation of the \lya forest and 21 cm intensity mapping. Mathematically,  
\begin{eqnarray}
P_{21,F} (\boldsymbol{k},z) = P_{21,F}^{\rm Fid.}(\boldsymbol{k},z) + P^{\rm Mem.}_{F}(\boldsymbol{k},z) + P^{\rm Mem.}_{21}(\boldsymbol{k},z) \, , 
\label{eq:power}
\end{eqnarray}
where the first term on the right-hand side corresponds to the conventional cross-correlation of the \lya forest and 21cm IM without reionization relics. The second term accounts for the memory of reionization present in the \lya forest, originating in the underdense regions. The final term represents the memory of reionization in \ion{H}{I} 21 cm IM, i.e. the memory of reionization sourced by overdense regions. The three terms are given by \citep{2017JCAP...04..001C,2019MNRAS.487.1047M,2023MNRAS.525.6036L} 
\begin{eqnarray}
&P^{\rm Fid.}_{21,F}  = b_F(z) b_{\ion{H}{I}}(z) [1 + \beta_F(z) \mu^2] [1 + \beta_{\ion{H}{I}}(z) \mu^2] P_m (k,z) \, ,  \\
&P^{{\rm Mem.}}_{F} = b_{\ion{H}{I}}(z) [1 + \beta_{\ion{H}{I}}(z) \mu^2] b_\Gamma(z) P_{m,\psi}(k,z) \, , \\
&P^{\rm Mem.}_{21} = b_F(z) [1 + \beta_F(z) \mu^2] P_{m,\Xi} (k,z) \, , 
\end{eqnarray}
where $b_F$ and $\beta_F$ are the usual flux bias and redshift space distortion parameter, respectively -- and similarly for 21 cm quantities. Moreover, $\mu$ is the angle with respect to the line of sight, $b_\Gamma$ is the radiation bias defined in \cite{2015JCAP...12..017A,2018MNRAS.474.2173H}, and $P_m$ is the linear matter power spectrum. 

Furthermore, $P_{m,\psi}$ and $P_{m,\Xi}$ are the cross-power spectrum of matter and change of transparency of the IGM due to the impact of reionization, and the cross-power spectrum of matter and change of neutral hydrogen density induced by the passage of reionization fronts, respectively. These cross-power spectra are defined as follows \citep{2019MNRAS.487.1047M,2023MNRAS.525.6036L}
\begin{eqnarray}
\label{eq:psi}
P_{m,\psi}(k,z_{\rm obs}) \!\!&=&\!\! - \int {\rm d}z \frac{\partial \psi(z,z_{\rm obs})}{\partial z} P_{m,x_{\rm HI}}(k,z)\frac{D(z_{\rm obs})}{D(z)}\, , \\ 
\label{eq:xi}
P_{m,\Xi}(k,z_{\rm obs}) \!\!&=&\!\! -\int {\rm d}z \frac{\partial \Xi(z,z_{\rm obs})}{\partial z} P_{m,x_{\rm HI}}(k,z) \frac{D(z_{\rm obs})}{D(z)}\, ,
\end{eqnarray}
where the integration covers the epoch of reionization, $P_{m,x_{\rm HI}}$ is our proxy for the correlation of matter and ionized bubble spatial distribution, which accounts for the patchy nature of reionization. $D$ is the growth factor, while $\psi$ denotes the transparency of the IGM, and $\Xi$ represents the modulation of the neutral hydrogen due to the passage of ionization fronts. They are given by 
\begin{eqnarray}
\label{eq:psi_ref}
&&\psi(z_{\rm re},z_{\rm obs}\,|\, \overline{z}_{\rm re}) = \Delta \ln \tau_1 =  \ln \left[ \frac{\tau_1(z_{\rm re},z_{\rm obs})}{\tau_1(\overline{z}_{\rm re},z_{\rm obs}) }\right]\, , \\
\label{eq:xi_ref}
&&\Xi(z_{\rm re},z_{\rm obs}\,|\,\overline{z}_{\rm re}) = \Delta \ln \rho_{\rm HI} = \ln \left[\frac{\rho_{\rm HI}({z_{\rm re},\,z_{\rm obs}})}{\rho_{\rm HI}({\overline{z}_{\rm re},\,z_{\rm obs}})}\right] \, ,
\end{eqnarray}
where $3.5 \leq z_{\rm obs} \leq 4$, $z_{\rm re}$ is the \textit{local} redshift of reionization, and $\overline{z}_{\rm re}$ serves as a reference redshift of reionization. The parameter $\tau_1$ is the optical depth that must be assigned in simulations to a patch of gas with mean density and temperature $T=10^4$ K in order for the mean transmitted flux to match observations, i.e. $\tau_1$ is a normalization factor that guarantees that an optical depth cube return sensible results by matching to the observations of \cite{2007MNRAS.382.1657K}. The transparency is defined as a relative measure, benchmarked against a fiducial scenario with $\overline{z}_{\rm re} = 8$. Conversely, $\Xi$ characterizes the response of halos with shallow potential wells to the passage of ionization fronts, modeling the perturbations in neutral hydrogen density induced by the response relative to those that occur in a fiducial local reionization scenario. We plot some of the ingredients of the reionization relics in Figure \ref{fig:fish-models} for reference. 

Similarly, the \lya flux and the 21 cm power spectra are respectively given by
\begin{eqnarray}
    \label{eq:f-power}
    &P_F^{\rm 3D} (\boldsymbol{k},z) = b_F^2 (1 + \beta_F \mu^2)^2 P_m + 2 b_F b_\Gamma (1 + \beta_F \mu^2) P_{m, \psi} \\
    &P_{21} (\boldsymbol{k},z) = b^2_{\ion{H}{I}} (1 + \beta_{\ion{H}{I}} \mu^2)^2 P_m + 2 b_{\ion{H}{I}} (1 + \beta_{\ion{H}{I}} \mu^2) P_{m, \Xi} 
\end{eqnarray}
where we have neglected terms with higher order in $\Xi$ and $\psi$. Likewise, we disregard non-linear corrections, justified by our focus on large scales.

Even though we extract $b_F$ and $\beta_F$ from \cite{2015JCAP...12..017A}, the maximum redshift in their tables is 3. Hence, for our $z > 3$ calculations, we include a redshift evolution factor $[(1 + z) / (1 + z_p) ]^{3.55}$ \citep{2013A&A...559A..85P} with pivot $z_p = 3$ to account for the evolution of the flux bias and redshift space distortion parameter \citep[see Eq.(10) and surrounding text of ][]{2024MNRAS.529.3666M}.

While both Eq.~(\ref{eq:psi}, \ref{eq:xi}) emerge from the IGM's response to inhomogeneous reionization, they represent distinct response mechanisms. Eq.~(\ref{eq:psi_ref}) covers the response to the local reionization process by underdense gas \citep{2018MNRAS.474.2173H}. Physically, it originates in minivoids, as underdense gas undergoes ultraviolet heating, shock heating, and compression during the reionization process. In contrast, Eq.~(\ref{eq:xi_ref}) captures the response to local reionization by denser regions, predominantly arising from minihalos. Essentially, this is the modulating effect on the number of baryons allowed inside a given halo following the \textit{thermal kick} delivered by an ionization front \citep{2022MNRAS.513..117L}. Note that both mechanisms differ in their origin on the small scales but are influenced equivalently by the patchy nature of reionization -- parametrized here as $P_{m,x_{\rm HI}}$ and present in both Eq.~(\ref{eq:psi}, \ref{eq:xi}).

To facilitate comparison between the different components of Eq.~(\ref{eq:power}), we have chosen to utilize the spherically-averaged power spectrum\footnote{This choice was also used by \cite{2015JCAP...03..034V} and accounts for the difficulty of having enough angular resolution to have a reliable measurement at different $\mu$-bins.}, i.e.
\begin{eqnarray}
P^{\rm Sph.}_{21, F} (k,z) = \frac{1}{2\pi} \int_0^{2\pi} d \phi \int_0^1 d\mu P_{21,F}(\boldsymbol{k},z)\, , 
\label{eq:sphP}
\end{eqnarray}
where the normalization factor already accounts for the extra factor of 2 due to the $\mu$-symmetry.

\begin{figure*}
    \centering
    \includegraphics[width=\linewidth]{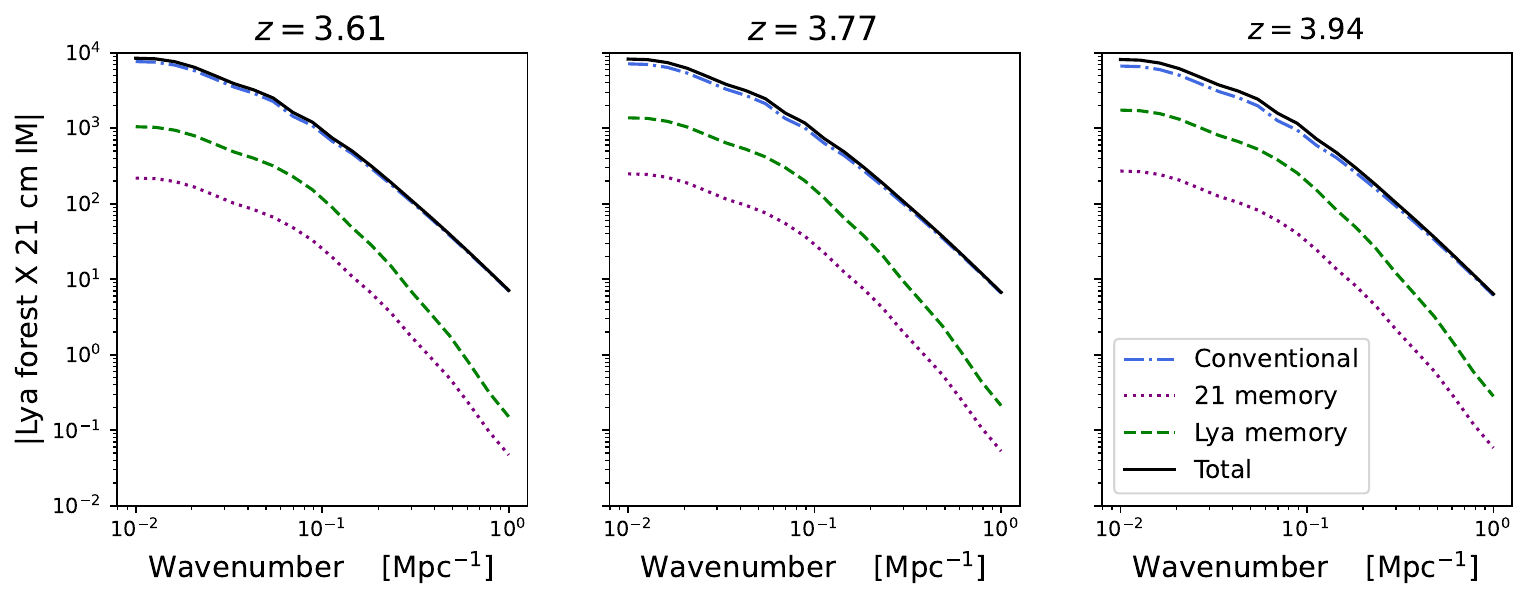}
    \caption{The magnitude of the spherically-averaged cross-power spectrum for the Lyman-alpha forest $\times$ \ion{H}{I} 21 cm IM cross-correlation is illustrated, encompassing its components, including the imprint of reionization in both probes. The blue dash-dotted line corresponds to the \textit{conventional} signal without any relics from cosmic reionization. The purple dotted and green dashed curves represent the memory of reionization in the Lyman-$\alpha$ forest and 21 cm power, respectively. We highlight that the memory of reionization in the 21 cm signal is negative, leading to a competition in which the \lya forest, which is positive, dominates. The influence of reionization becomes more pronounced at higher redshifts and larger scales.}
    \label{fig:power}
\end{figure*}

We plot Eq.~(\ref{eq:sphP}) in Figure \ref{fig:power} at a few selected redshifts. The impact of reionization is more pronounced at higher redshifts, introducing a competition between the 21 cm effect, arising from the modulation of baryons due to reionization, and the \lya forest effect, originating in underdense regions. The former aims to diminish the signal's amplitude while the latter enhances it. Furthermore, the more substantial (in absolute value) impact of the \lya sourced term compared to the \ion{H}{I} 21 cm term is anticipated due to its greater influence on voids, evident in the bimodal temperature-density relation. See Figure 3 to 5 in \citealt{2020MNRAS.499.1640M} versus baryon modulation displayed in Figure 9 of \citealt{2023MNRAS.525.6036L}. The modulation effect of reionization on the baryon content of halos contributes up to approximately 25\% of the overall 21 cm intensity mapping signal at $z = 5.5 $, while the impact of inhomogeneous reionization on the \lya forest accounts for roughly 60\% of the \lya power spectrum at $z = 4$. While our focus is on the $3.5<z<4.$ redshift range, we expect this qualitative trend -- a more pronounced reionization imprint on the \lya forest -- to persist up to higher redshifts ($4 \leq z \lesssim 6$) especially given that the impact of self-shielding in minihalos will also enhance the \lya power spectrum at large scales \citep{2024ApJ...969...46P} because the surviving neutral hydrogen protected in those dense gas regions will lead to a strong absorption at high redshifts, boosting the power spectrum at large scales, which aligns with the effect of inhomogeneous reionization. Regarding the wavenumber trend, the larger deviation at large scales compared to the conventional term is explained by the coupling to the reionization bubble scale in the cross-correlation of the matter and neutral hydrogen fraction field, $P_{m, x_{\ion{H}{I}}}$.

\section{Simulations}
\label{sec:sims}
To accurately model the reionization process, it is imperative to probe under the Jeans length prior to the passage of an ionization front ($\sim 100$ kpc). Failure to do so would result in the loss of the ability to track the response by the small-scale structure due to the significant wiping out of these structures caused by the increasing Jeans scale throughout reionization \citep{2018MNRAS.474.2173H}. Besides, reionization occurs in an inhomogeneous way. Thus, to secure enough statistical power, there is a need to simulate large comoving volumes (a few hundred Mpc) with substantial ionized bubbles each with a radius of a few Mpc. However, the dynamical range required for these considerations is too large to accomplish with reasonable computational resources using a single simulation. 

We adopt the methodology introduced by \cite{2019MNRAS.487.1047M,2023MNRAS.525.6036L} and employ a hybrid approach to compute Eq.~(\ref{eq:psi},\ref{eq:xi}). This approach relies on seminumerical simulations using {\sc 21cmFAST}  \citep{2011MNRAS.411..955M,2020JOSS....5.2582M}, which track the patchy nature of reionization within a 400 Mpc box. In addition, we leverage high-mass resolution small-box ($L = 1152 h^{-1}$ ckpc) simulations run using a modified version of {\sc Gadget-2} \citep{2005MNRAS.364.1105S,2018MNRAS.474.2173H} to monitor the response to ionization fronts in both dense (crucial for \ion{H}{I} 21 cm) and underdense (Ly$\alpha$ forest) regions. This hybrid simulation strategy allows us to capture the impact of reionization across a range of scales and redshifts with minimal trade-offs. Naturally, this hybrid methodology will not be sensitive to higher-order correlations between the reionization field and the small-scale density field \citep{2019MNRAS.487.1047M}. 

The small-box simulations include adiabatic expansion (including Hubble expansion), shock heating, Compton heating, and cooling for neutral gas (accounting for residual ionization). In addition, we incorporate the long-lasting impact of X-ray preheating in the post-reionization IGM \citep{2024MNRAS.529.3666M}. For ionized gas, we also include Compton cooling, \ion{He}{II} cooling, recombination cooling, photoionization heating, and free-free cooling. These simulations have been described -- and tested -- in detail in \cite{2018MNRAS.474.2173H,2022MNRAS.513..117L}.
 
To surpass the limitation on the smallest wavenumber imposed by the size of our {\sc 21cmFAST} boxes, we utilize a simple liner bias to estimate the effect of reionization at $k \lesssim 0.06$ Mpc$^{-1}$ as follows
\begin{eqnarray}
    P_{m, X} (k,z) = \frac{P_{m,X}(k_{\rm cut} \approx 0.06,z)}{P_m (k_{\rm cut},z)} P_m(k,z) \, ,
    \label{eq:cut}
\end{eqnarray}
where $X = \{\psi, \Xi\}$. This approximation should be sufficient for larger scales than the ionized bubble scale. Note that this restriction is due to the box side of our large-box ($400$ Mpc) simulations and it is a self-imposed restriction based on available computational resources. Given the size of our simulation box, the largest scale we can probe is $k_{\rm min} \approx 0.02$ Mpc$^{-1}$. However, the number of modes in the first few $k$-bins would be too small to provide meaningful statistical power. Thus, we choose $k_{\rm cut} = 0.06$ Mpc$^{-1}$ as our cutoff.

\section{Telescopes}
\label{sec:tele}
To measure the cross-correlation between 21 cm and the \lya forest, we require a radio interferometer and a spectroscopic telescope with some non-negligible overlap in sky coverage and redshift. Unfortunately, our choice for estimating the SNR is not straightforward due to the timeline of the Dark Energy Spectroscopic Instrument \citep[DESI; ][]{2022AJ....164..207D} versus the timeline of the Square Kilometer array \citep[SKA; ][]{2019arXiv191212699B}. While DESI may potentially extend its operational life as DESI-II, allowing for direct operational overlap with SKA, the exact timeline overlap remains uncertain.  

Concurrently, upcoming spectroscopic instruments such as MegaMapper \citep{2019BAAS...51g.229S}, the MUltiplexed Survey Telescope \citep[MUST;][]{2024MNRAS.530.1235Z}, and MaunaKea Spectroscopic Explorer \citep[MSE;][]{2019arXiv190303158P}, categorized as Stage V spectroscopic instruments, are anticipated to be operational around the time when 21 cm radio interferometers come online. However, we underscore that the detailed designs of these instruments are currently in the developmental phase. Navigating the evolving landscape of these instrument timelines and designs will be vital to optimizing an observational program capable of measuring the cross-correlation between 21 cm and the \lya forest. 

Given these ``instrumental'' constraints, we have opted to explore several distinct instrumentation scenarios.
\begin{itemize}
    \item SKA1-Low $\times$ DESI-like spectroscopic telescope (referred to as SKA in relevant figures).
    \item SKA1-Low $\times$ DESI-MUST-like hybrid (referred to as MUST). 
    \item PUMA-like $\times$ DESI-MUST-like hybrid (referred to as PUMA).
\end{itemize}
Across all scenarios, we maintain a redshift coverage of $3.5 \leq z \leq 4$ and consider two sky coverage overlaps -- 100 and 1000 square degrees. Likewise, we vary the integration times, considering both 100 and 1000 hours, which will mainly affect the radio observations as described below.

Our base reference scenario involves the SKA1-Low $\times$ DESI-like pair, with 100 hours of integration time and a 100-square-degree overlap between the instruments. This reference scenario assumes a Planck-like reionization timeline, which may be underestimating the impact of reionization since Planck uses a hyperbolic tangent as a sigmoid to model reionization that proves a poor fit of astrophysical constraints on the timeline of reionization \citep[see Figure 4 of ][]{2024arXiv240513680M}.

\subsection{SKA1-Low}
\label{ssec:ska}
SKA1-Low, our baseline 21 cm instrument, operates as an aperture array radio interferometer. We consider only the dense core of the array, which is most sensitive to the 21 cm power spectrum. Do note that the outer stations in the array are crucial for calibration and foreground removal purposes. The SKA1-Low, upon completion, offers versatility in supporting various observing modes. Previous works have often focused on observing mode 1, characterized by a dense core comprising 224 stations of 40 meters in diameter, each containing 256 dipole antennas \citep{2020PASA...37....7S,2022ApJ...933..236Z}. These core stations are distributed over a diameter of approximately one kilometer.

Nevertheless, as pointed out in \href{https://www.skao.int/sites/default/files/documents/d17-SKA-TEL-SKO-0000557_01_-DesignConstraints-1.pdf}{SKA1-LOW Configuration - Constraints \& Performance Analysis} \S5.4.3, power spectrum measurements are better served by ``substations'' with 10 m of diameter. Here we chose to consider only the observing mode 4, as detailed in Table 2 of \href{https://www.skao.int/sites/default/files/documents/d17-SKA-TEL-SKO-0000557_01_-DesignConstraints-1.pdf}{SKA1-LOW Configuration - Constraints \& Performance Analysis}. This mode, supported by the instrument's correlator, boasts six times more correlatable elements (substations) at the expense of a reduced bandwidth (8.4 MHz compared to 300 for mode 1).

To quantify the unique baselines covered by the dense core, we utilize the number density of baselines $n_b$ from \cite{2015JCAP...03..034V} and normalize it to recover the total amount of distinct baselines covered by the dense core, as outlined in \S2 of \cite{2023MNRAS.525.6036L}. 

The system temperature -- the sum of the sky temperature, the ground reflections, and the temperature of the receiver -- is given by
\begin{eqnarray}
T_{\rm sys} (\nu) = 60  \left(\frac{300 \ \textup{MHz}}{\nu} \right)^{2.55} \times 1.1 + 40 \ \ \ \  \textup{[K]}, 
\end{eqnarray}
while the effective collecting area per station $A_e$ is given by \citep{2020PASA...37....7S}
\begin{eqnarray}
    \label{eq:sk_A}
    A_e (\nu) =  A_{e, \rm{crit}} \times \begin{cases} \left(\frac{\nu_{\rm crit}}{\nu}\right)^2, & \nu > \nu_{\rm crit} \\
    1, & \nu \leq \nu_{\rm crit}
    \end{cases}~,
\end{eqnarray}
where $\nu_{\rm crit} = 110$ MHz and $A_{e, \rm{crit}}$ is the collecting area in m$^2$ for the 256 dipole antennas of 3.2 m$^2$ each for observing mode 1. Thus, we approximate the effective area for observing mode 4 by decreasing Eq.~(\ref{eq:sk_A}) by a factor of 6.

Moreover, the field of view is given by 
\begin{eqnarray}
    \label{eq:fov}
    {\rm FoV} (\lambda) = \left(\frac{\lambda}{\sqrt{0.7} D_{\rm phys}}\right)^2 \, ,
\end{eqnarray}
where $D_{\rm phys}$ is the diameter of the correlatable element and $\lambda$ is the observing wavelength.

\subsection{PUMA}
\label{ssec:puma}
The Packed Ultra-wideband Mapping Array design \citep[PUMA; ][]{2018arXiv181009572C,2019BAAS...51g..53S} -- \textit{currently} -- consists of 32000 antennas distributed in a hexagonal-close packing in a compact circle of roughly 1.5 km diameter. Each antenna has a diameter of 6 meters. We take the baseline number density $n_{\rm b}$ directly from Appendix D of \cite{2018arXiv181009572C}. 

For PUMA, the system temperature is given by
\begin{eqnarray}
    T_{\rm sys}(\nu) = 25 \left(\frac{400 \ \textup{MHz}}{\nu} \right)^{2.75} + 2.7 + \frac{300}{9}+ \frac{50}{0.81} \ \ \ \ \textup{[K]}.
\end{eqnarray}

The effective collecting area per antenna is simply the effective area of the antenna dish, i.e. $\pi (\sqrt{0.7} D_{\rm phys} / 2)^2$. Simultaneously, the field of view is given by Eq.~(\ref{eq:fov}) but with $D_{\rm phys} = 6$ m. We set the bandwidth to 8.4 MHz (same as SKA1-Low observing mode 4).

We consider PUMA as an optimistic/futuristic 21 cm telescope for our purposes, essentially our best benchmark.

\subsection{DESI-like}
\label{ssec:desi}
We assume a DESI-like spectrograph as our baseline \lya forest instrument, which is likely a conservative estimate given the projected timeline of SKA. To gauge the performance of the final -- 5 years -- DESI data, we use the Quasar Luminosity Function (QLF) to quantify the expected number of \lya quasars observable with DESI \citep{2013A&A...551A..29P,2020RNAAS...4..179Y} and the spectrograph performance of DESI -- see Eq.~(\ref{eq:lya-tot}). Detailed information about the DESI instrument is available in \cite{2022AJ....164..207D}, the QSO and \lya QSO target selection can be found in \cite{2023ApJ...944..107C}, and the spectroscopic pipeline is described in \cite{2023AJ....165..144G}.

\subsection{DESI-MUST-like hybrid}
\label{ssec:must}
We refine our initially conservative DESI-like scenario by enhancing its performance to align more closely with the standards expected for Stage V spectroscopic instruments. Pragmatically, this improvement involves adjusting the aliasing term in the covariance computation, effectively augmenting the effective density of lines of sight beyond what is anticipated with our original DESI specs. In summary, we implement an optimistic factor of 3 reduction in both the second and third terms of Eq.~(\ref{eq:lya-tot}). Given that MUST will likely be the World's first Stage V spectroscopic instrument\footnote{MUST's first light is currently planned for 2029, see \url{https://must.astro.tsinghua.edu.cn/en} for more details.}, we decided to refer to this scenario as MUST.

\section{Results}
\label{sec:results}
In the redshift range of interest, the cross-correlation of \lya $\times$ 21 cm IM presents a pragmatically more accessible measurement compared to the auto-correlation analyses of both the \lya forest or of the \ion{H}{I} 21 cm field. This disparity arises from distinct challenges associated with each observable. For the \lya forest, the QLF peaks around $z \sim 2$, resulting in a substantial reduction in the number density of quasars at $z \sim 4$. Conversely, for 21 cm intensity mapping, the signal strength is significantly weaker than the foregrounds by several orders of magnitude. Notably, these challenges are uncorrelated, rendering the cross-correlation less susceptible to their impacts \citep{2017JCAP...04..001C}. Since this is the first study of the effects of the relics from reionization in this observable, we will not consider foreground contamination nor several \lya forest systematics, like continuum fitting \citep{2023ApJS..269....4S}, spectra with broad absorption lines \citep{2023arXiv230903434F}, UV clustering \citep{2023MNRAS.520..948L}, spectra with damped Lyman-$\alpha$ systems \citep{2022ApJS..259...28W}, and several other astrophysical and instrumental systematics. Note that this simpler approach is often used in other studies of this cross-correlation \citep[e.g. ][]{2011MNRAS.410.1130G,2019JCAP...12..058S,2017JCAP...04..001C} at other redshifts.

\subsection{Impact on SNR}
\label{ssec:snr-just}

Under these considerations, we can compute the errors on the observables following \cite{2017JCAP...04..001C,2015JCAP...03..034V}
\begin{eqnarray}
    \label{eq:si-21}
    \sigma^2[P_{21}(\boldsymbol{k},z)] &=& \left(P^{\rm T}_{21} (\boldsymbol{k},z) \right)^2 \, , \\ \label{eq:si-f}
    \sigma^2[P_F(\boldsymbol{k},z)] &=& \left(P^{\rm T}_{F} (\boldsymbol{k},z) \right)^2 \, \, \textup{and} \\ \label{eq:si-x}
    \sigma^2[P_{21, F}(\boldsymbol{k},z)] &=& \frac{1}{2} \left(P^2_{21,F}  + \sigma [P_{21}] \sigma [P_F]    \right) \, .
\end{eqnarray}
For 21 cm, $P^{\rm Tot.}$ has contributions from cosmic variance and thermal noise, i.e. 
\begin{eqnarray}
    \label{eq:21-tot}
    P^{\rm T}_{21}(\boldsymbol{k}, z) &=& P_{\ion{H}{I}} + T^2_{\rm sys}(z) \chi^2(z) \lambda (z) \frac{1 + z}{H(z)} \left(\frac{\lambda^2 (z)}{A_e}\right)^2 \left(\frac{S_{\rm area}}{{\rm FoV}(z)}\right) \nonumber \\
    & & \times \left(\frac{1}{N_{\rm pol} t_{\rm int} n_{\rm b}(u = k_{\perp} \chi (z)/ 2\pi)}\right)  \, ,
\end{eqnarray}
where $\chi$ is the comoving distance, $H$ is the Hubble parameter, and $\lambda = 21 \ {\rm cm} \, / (1 + z)$. Moreover, $T_{\rm sys}$ stands for the system temperature of the radio interferometer, $N_{\rm pol} = 2$ is the number of polarizations, $A_e$ the effective area of the correlatable unit (dishes or stations), $S_{\rm area}$ is the survey area, FoV is the field of view, $n_{\rm b}$ quantifies the number density of baselines in the $uv$-plane as a function of wavenumber, and $t_{\rm int}$ corresponds to the integration time. The shot noise is subdominant for the redshift range of interest here \citep{2017MNRAS.471.1788C}, thus we do not include it in Eq.~(\ref{eq:21-tot}).

In contrast, for the \lya forest we have \citep{2007PhRvD..76f3009M,2014JCAP...05..023F,2021MNRAS.508.1262M}
\begin{eqnarray}
    \label{eq:lya-tot}
    P^{\rm T}_{F} (\boldsymbol{k},z) = P_F^{\rm 3D}(\boldsymbol{k},z) + P_F^{\rm 1D}(k_\parallel,z)P^{\rm 2D}_w(z) + P_N^{\rm eff}(z) \, . 
\end{eqnarray}
The first term corresponds to cosmic variance, i.e. it comes from $\langle\delta_F \delta_F^*\rangle$, while the second term represents the aliasing noise due to the sparse sampling of quasars. The third term describes the effective noise due to the spectrograph performance. Note that at high redshifts ($z \gtrsim 3.2$), the aliasing term tends to dominate Eq.~(\ref{eq:lya-tot}) since the limited number of quasars becomes more restrictive \citep[see Figure 6 of][]{2021MNRAS.508.1262M}. Naturally, this trend reduces the expected signal-to-noise ratio (SNR) of \lya forest surveys attempting a measurement of the 3D flux power spectrum at high redshifts. 

Given Eq.~(\ref{eq:si-21}, \ref{eq:si-f}, or \ref{eq:si-x}), the SNR for $P_i$ can be written as
\begin{eqnarray}
    \label{eq:snr}
    {\rm SNR}^2_i = N_k \frac{P^2_i}{\sigma^2[P_i]} = \frac{V_{\rm Survey} k^3 \epsilon d\mu}{4 \pi^2} \frac{P^2_i}{\sigma^2[P_i]} \, ,   
\end{eqnarray}
where $\epsilon = dk/k$ and $V_{\rm Survey}$ is the -- overlap -- volume between the radio interferometer and the spectrograph. We consider three redshift bins centered\footnote{The center of the bins were chosen to coincide with the last three bins used in \cite{2021MNRAS.508.1262M}'s forecast.} at $z_{\rm c} = [3.61, 3.77, 3.94]$ with widths defined by the bandwidth of the radio interferometer, i.e. $\Delta \nu = 8$ MHz. Consequently, our survey volume for a redshift bin centered at $z_{\rm c}$ is given by
\begin{eqnarray}
    \label{eq:v_sur}
    V_{\rm Survey} (z_{\rm c}) = \frac{4 \pi}{3} f_{\rm sky} [\chi(z_{\rm max})^3 - \chi(z_{\rm min})^3)] \,
\end{eqnarray}
with $\chi$ as the comoving distance. Note that $f_{\rm sky}$, the sky coverage, depends on the area of overlap. 

In what follows, we focus on the spherical-averaged power spectrum, i.e. Eq.~(\ref{eq:sphP}) and analogs, to compute the appropriate SNR for \lya $\times$ 21 cm, \lya forest, and 21 cm intensity mapping. 

\begin{figure*}
    \centering
    \includegraphics[width=\linewidth]{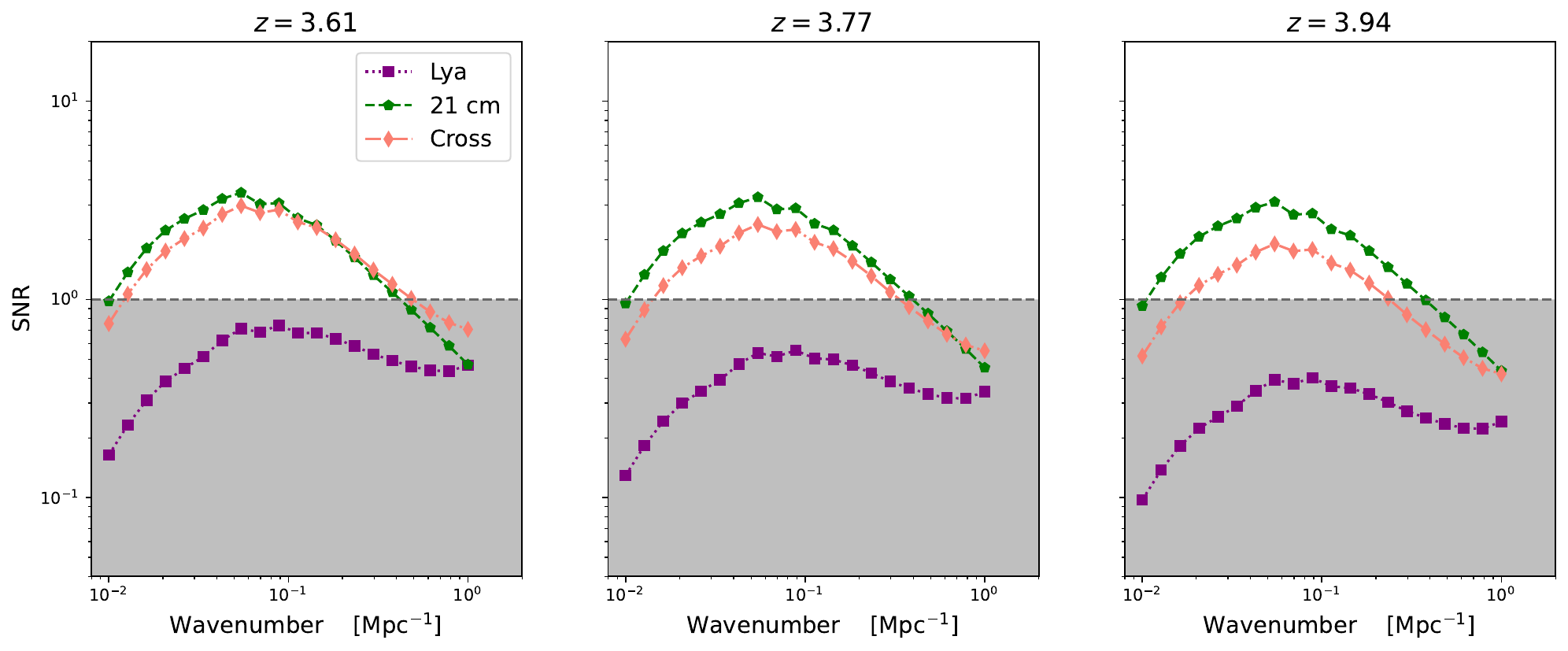}
    \caption{The SNR for the baseline configuration of the \lya forest (purple squares), 21 cm intensity mapping (green pentagons), and Ly$\alpha$ $\times$ 21 cm (orange diamonds) is presented as a function of wavenumber in our baseline scenario (see text for details). The three panels correspond to different redshift bins. The cross-correlation is, naturally, the more pragmatic measurement since the 21 cm signal would be severely impacted by foregrounds. Notably, the SNR for the forest is somewhat impeded by the chosen survey volume, although the drop in the number density of quasars leads to difficulties at high redshifts regardless of sensible choices for survey volume. The gray-shaded region corresponds to SNR $\leq 1$. }
    \label{fig:SN_just}
\end{figure*}

We illustrate the observability of the cross-correlation compared to the auto-correlation of \lya and 21 cm in Figure \ref{fig:SN_just} for our base scenario (SKA1-Low $\times$ DESI-like, $t_{\rm int} = 100$ h, $A_{\rm sky} = 100$ sq. deg. and Planck's reionization timeline). We highlight that all of the curves here include the long-lasting relics from cosmic reionization, i.e. Eqs.~(\ref{eq:psi}, \ref{eq:xi}). Unsurprisingly, the SNR for the \lya forest is less than unity because of the drop in available line-of-sight at these redshifts and because of the survey volume. In principle, it is necessary to observe more quasars to reduce the mean separation between forests to make the 3D flux power spectrum observation feasible. Although Figure \ref{fig:SN_just} demonstrates that the 21 cm signal has a larger SNR, this is only true in the absence of foregrounds, which will severely bury the auto-power spectrum. In contrast, the cross-correlation is more robust against foreground contamination \citep{2007ApJ...660.1030F, 2018JCAP...05..051S, 2021ApJ...909...51Z} and it is already competitive with the 21 cm auto-correlation with the baseline instrumental setup. 

Figure \ref{fig:SN_just} provides a compelling rationale for the significance of the \lya $\times$ 21 cm cross-correlation, particularly in the presence of foregrounds. Detection in the cross-correlation can guarantee the cosmological nature of the 21 cm signal. Besides, the SNR is similar to the auto-correlation at the ionized bubble scale (and for our first redshift bin). This is consistent with the findings of \cite{2017JCAP...04..001C} at a lower redshift ($z = 2.4$). However, we underscore that in the absence of reionization relics, the cross-correlation will not have a similar SNR to that of the auto 21 cm SNR in the lowest redshift bin considered here. The boost provided by the memory of reionization is crucial to be competitive, yet it is not sufficient at higher redshift bins because of a significant rise in aliasing noise -- $P_w^{\rm 2D}$ in Eq.~(\ref{eq:lya-tot}) -- at those redshifts. 

Furthermore, akin to the findings in \cite{2017JCAP...04..001C}, the cross-correlation can surpass the SNR of the 21 cm auto-correlation at smaller scales. However, this trend is obscured at high redshifts due to the poor sampling of quasars. 

Having established the interest in the cross-correlation of the \lya forest and 21 cm IM at the redshift range of interest ($3.5 \leq z \leq 4$), we now ponder the impact of the survey strategy.

\subsection{Importance of survey design}
\label{ssec:snr-survey}
Our main objective here is to underscore the importance of a non-negligible overlap between spectroscopic instruments and radio interferometers. Given that SKA1-Low is under development in the southern hemisphere and DESI -- also likely DESI II -- is in the northern hemisphere, we investigate the potential advantages and insights that could be derived from this cross-correlation with diverse instrumental configurations.

In order to reduce the noise in the radio interferometer, a widely employed strategy is to consider an increase in the integration time since the thermal noise is inversely proportional to $t_{\rm int}$. Hence, we explore the impact of a tenfold increase in $t_{\rm int}$ across various instrumental configurations -- specifically SKA1-Low $\times$ DESI-like, SKA1-Low $\times$ DESI-MUST-like hybrid, and PUMA-like $\times$ DESI-MUST-like hybrid.

We assume that the \lya forest observations will be completed regardless of the integration time; nevertheless, we reward the longer observation time with a factor of two reduction in the spectrograph effective noise $P_N^{\rm eff}$. In other words, this implies that the extended observation time allows for more exposures. Note that we do not modify the aliasing error -- second term in Eq.~(\ref{eq:lya-tot}) that significantly dominates the error budget -- even when considering additional observation time.

\begin{figure*}
    \centering
    \includegraphics[width=\linewidth]{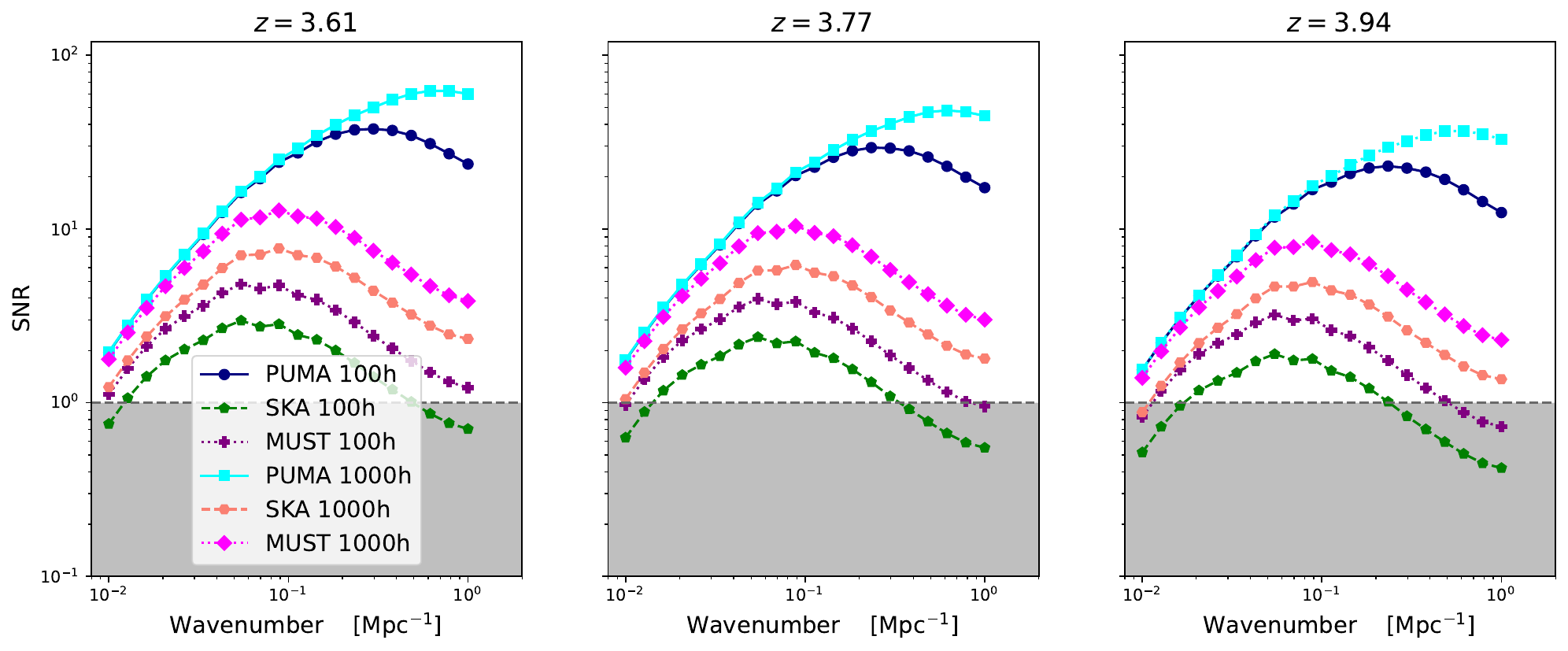}
    \caption{SNR for the Ly$\alpha$ forest $\times$ 21 cm intensity mapping accounting for variations in the integration time of the radio telescopes. We assume that the \lya forest survey is also completed during that integration time and reward the additional integration time (see text for details). Shown are the SNR for: PUMA $\times$ MUST 100 hours (blue circles) and 1000 hours (cyan squares); SKA1-Low $\times$ DESI 100 hours (green pentagons) and 1000 hours (orange octagons); SKA1-Low  $\times$ MUST 100 hours (purple crosses) and 1000 hours (magenta diamonds). Increasing the total integration time allows for a better measurement in all configurations. } 
    \label{fig:SN_tint}
\end{figure*}

Figure \ref{fig:SN_tint} showcases the dependence of the SNR for different integration times and telescope pairs. Unsurprisingly, there is a consistent enhancement in SNR across all redshifts and configurations. In particular, the SKA1-Low configurations benefit the most from this strategy, manifesting improvements of $\approx 2.6$ in overall SNR (see Table \ref{tab:snr}). Intriguingly, the PUMA-like $\times$ DESI-MUST-like hybrid exhibits negligible improvement at large scales. 

From Eq.~(\ref{eq:21-tot}), the thermal noise of the radio interferometer is inversely proportional to the total integration time. The reason Figure \ref{fig:SN_tint} shows a larger impact for the pairs using SKA1-Low is because the radio telescope dominates the error budget for those setups, i.e. it is larger than the cosmic variance contribution and $\sigma[P_{\ion{H}{I}}] > \sigma[P_F]$. This is also why the change from DESI-like to DESI-MUST-hybrid does not result in a major improvement and largely conserves the increase in SNR that was present in the 100-hour baseline scenario. In contrast, for the PUMA $\times$ MUST setup, the \lya forest survey now dominates the error, gains are then possible at larger wavenumbers given the large drop at large scales (as can be seen in Figure \ref{fig:SN_just}). We also highlight that the number density of baselines also restricts the shape of the SNR at these scales, which is why the SKA1-Low $\times$ MUST can be unexpectedly competitive with PUMA $\times$ MUST's SNR at some wavenumbers for a 1000 hours of integration time.    

It is unclear whether major projects like SKA or a Stage V spectroscopic instrument could allocate 1000 hours/many exposures to observe the same overlapping area, given the promising science programs they aim to address. Nevertheless, it is evident that with such a time investment the community could expedite the progress in measurements of the cross-correlation by a decade, achieving observations competitive with what may be available with Stage II 21 cm radio telescopes -- at least at large scales. We acknowledge, however, that this is a significant time commitment. 

\begin{figure*}
    \centering
    \includegraphics[width=\linewidth]{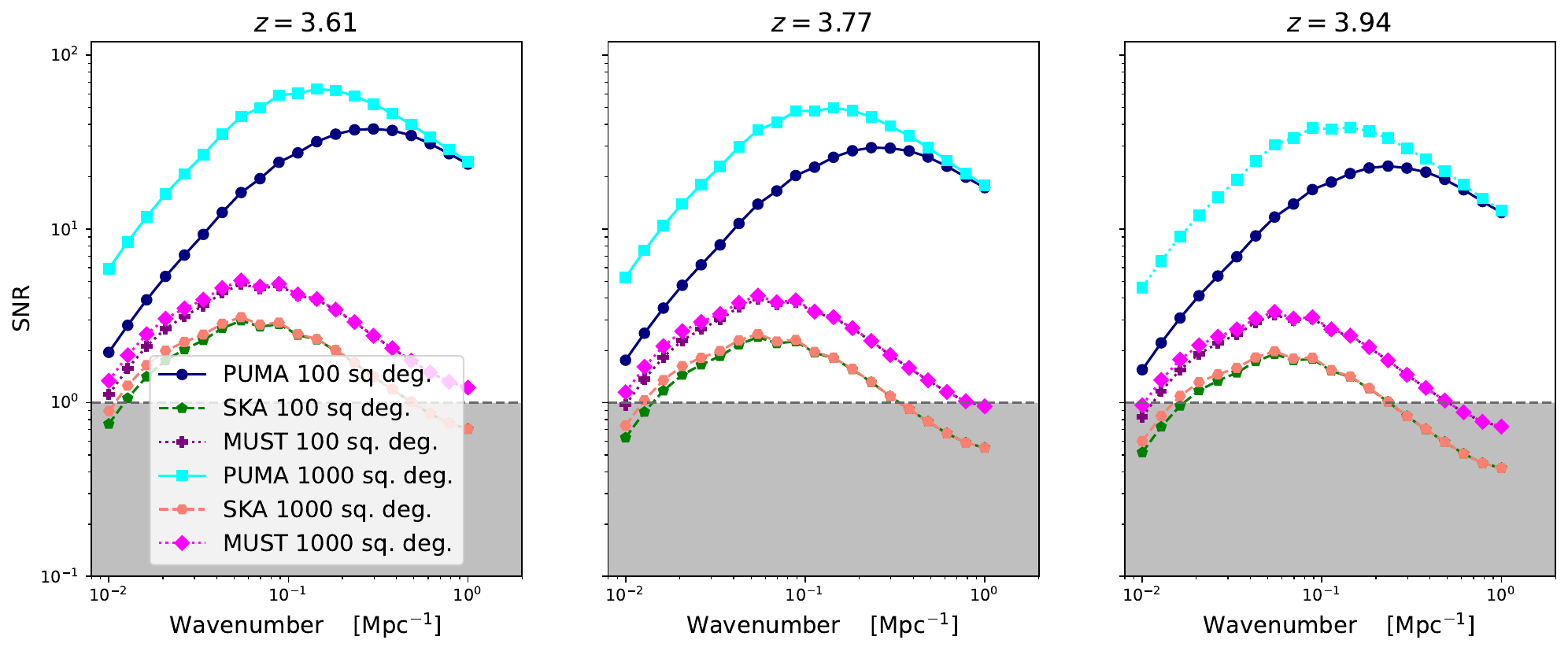}
    \caption{Similar to Figure \ref{fig:SN_tint}, but modifying the survey area. Notably, increasing the survey area does little to improve the SNR for the configurations with SKA1-Low since the error is dominated by the interferometric instrument and the thermal noise increases with survey area -- see Eq.~(\ref{eq:21-tot}).}
    \label{fig:SN_sky}
\end{figure*}

Naively, one might expect that expanding the overlapping area would perhaps be a more straightforward enterprise, especially for spectroscopic instruments located in the southern hemisphere like the 4-metre Multi-Object Spectroscopic Telescope \citep[4MOST;][]{2019Msngr.175...50R}. However, it is important to note that increasing the survey area produces a larger thermal noise -- see Eq.~(\ref{eq:21-tot}). Conversely, the number of modes $N_k$ is directly proportional to the survey area. Thus, as long as the thermal noise is manageable, an increase in the survey area will improve SNR due to the additional modes. For the PUMA setup, where \lya dominates the error budget at large scales, one could anticipate that the increase in modes may give rise to an enhancement in SNR up to a large $k$ value. In contrast, the impact of this choice is expected to generate limited gains for the SKA and MUST configurations due to the dominance of the 21 cm error, particularly with a tenfold increase in thermal noise. 

We confirm these expectations in Figure \ref{fig:SN_sky} by augmenting the area of overlap by a factor of ten from the baseline scenario (from 100 to 1000 square degrees). 

The interpretation of this finding is convoluted. Initially, it may be more appealing to allocate 100 hours of a telescope's time to an observation spanning different pointings. Unfortunately, this approach does not lead to a significant enhancement in observations for the late 2020s configurations. In terms of the overall SNR, aggregated across the $k$-bins, there is only a very modest increase of approximately $1.05$ compared to the baseline scenario.

Moreover, if radio telescopes happen to be located in the global south and spectroscopic instruments keep the trend of being located in the northern hemisphere, then the overlap may be small to begin with and hence preference would be given to securing a small overlap in the survey footprints. To our knowledge the possible location of PUMA is not decided yet, consequently, our results advocate for the consideration of overlap with Stage V spectroscopic instruments. Furthermore, our results also showcase that a larger overlap can produce significant gains for a PUMA cross-correlation at large scales. For PUMA, the overall SNR augments by a factor of $1.7$ when enlarging the survey area by a factor of 10. Besides, this enhancement is limited by the assumption of Stage V instrument performance. In principle, PUMA may instead share the skies with potential Stage VI spectroscopic instruments. 

While we have only scratched the surface regarding the implications of survey strategy for the \lya $\times$ \ion{H}{I} 21 cm cross-correlation at $3.5 \leq z \leq 4$, we deem further exploration of the instrumental setup beyond the scope of this work. This study focuses on demonstrating the impact of cosmic reionization in the cross-correlation during the post-reionization era. Therefore, it would be remiss of us not to address the dependence of the signal on the uncertain timeline of reionization \citep{2021MNRAS.508.1262M,2023ApJ...942...59J}.

\subsection{Dependence on reionization history}
\label{ssec:reio}
The timeline of reionization remains uncertain, although abundant direct detection of the ionizing sources is becoming possible thanks to the James Webb Space Telescope (JWST) \citep{2006SSRv..123..485G}. In fact, JWST has already started to revolutionize our understanding of galaxy formation at the epoch of reionization (EoR) redshifts \citep[e.g. ][]{2022arXiv221001777B,2023MNRAS.518.6011D,2023MNRAS.519.1201A,2023MNRAS.518.4755A}. Nevertheless, even though perfect knowledge of the reionization timeline would help to model the EoR, it would still be non-trivial to translate these constraints into a single mapping of reionization astrophysics. 

Consequently, it is intriguing to ponder the impact of different reionization scenarios in our SNR. We limit our study of reionization scenarios to three distinct reionization timelines. Our fiducial case aligns with Planck's timeline of reionization $z_{\rm mid} = z(\overline{x}_{\ion{H}{I}} = 0.5) \approx 7.7$, besides its duration, defined as $\Delta z = z(\overline{x}_{\ion{H}{I}} = 0.10) - z(\overline{x}_{\ion{H}{I}} = 0.90)$ is $3.83$. Furthermore, we complement this model with scenarios representing later and earlier reionization scenarios, having midpoints at $z_{\rm mid} = 6.88$ and $8.41$, and durations of $\Delta z = 4.15$ and $3.59$, respectively. Note that these additional models are roughly consistent with the 1$\sigma$ error reported by \cite{2020A&A...641A...6P} but we caution that the best fit from Planck is not able to reproduce astrophysical constraints in a satisfactory way -- see Figure 4 of \cite{2024arXiv240513680M} -- due to the use of the hyperbolic tangent reionization model \citep{2008PhRvD..78b3002L}. A more adequate choice would be to consider the range of reionization profiles allowed in the Gompertz model of reionization. Its best fit to CMB data, in conjunction with astrophysical constraints, indicates a midpoint of reionization at $\approx 7$ \citep{2024arXiv240513680M}, more closely aligned with our late reionization scenario. 

We obtain our three reionization models by modifying the ionization efficiency $\zeta$ in {\sc 21cmFAST}. This parameter governs the overall timeline of reionization, although there are degeneracies \citep{2014ApJ...782...66P,2019MNRAS.484..933P,2021MNRAS.508.1262M}. Physically, it quantifies the ability of photons to escape their parent galaxies and reach the intergalactic medium \citep{2018PhR...780....1D}.

\begin{figure*}
    \centering
    \includegraphics[width=\linewidth]{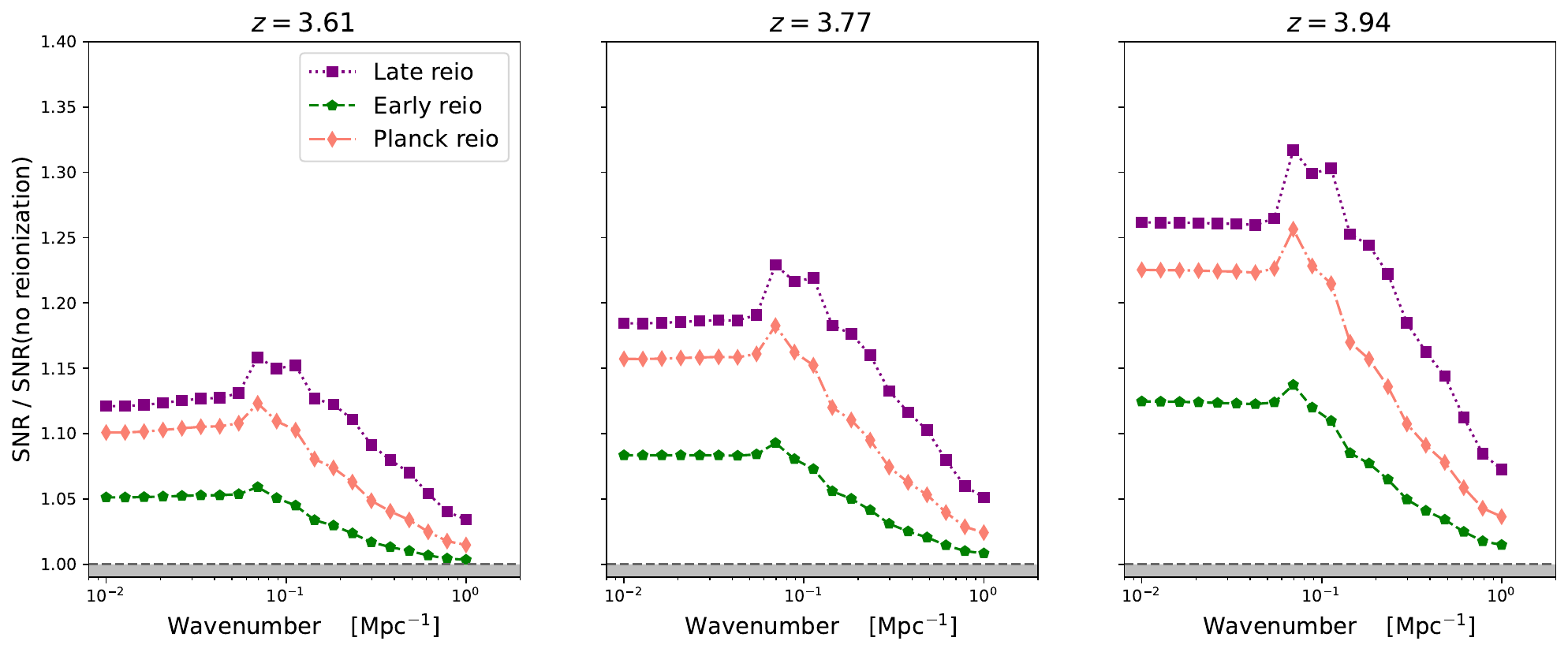}
    \caption{Impact of reionization history in the signal-to-noise ratio of the cross-correlation assuming the SKA1-Low $\times$ DESI baseline configuration, i.e. $t_{\rm int} = 100$ h and $S_{\rm survey} = 100 $ deg$^2$. Shown are the SNR for a late reionization (purple squares), a Planck-like reionization (orange diamonds), and an early reionization scenario (green pentagons). The behavior at small $k$ is driven by Eq.~(\ref{eq:cut}).}
    \label{fig:SN_reio}
\end{figure*}

In Figure \ref{fig:SN_reio}, we illustrate the effect of including the memory of reionization in $P_{21,F}$ compared to neglecting its existence. Naturally, the influence of the remnants of reionization is more pronounced in the highest redshift bin and for the delayed reionization scenario. As time progresses, gas dissipates the additional injected energy during reionization, resulting in a lesser effect in the \lya forest. Meanwhile, the modulation of the baryons in shallow potential wells becomes more subdued due to the growth of affected galaxies, as a result, the significance of the memory of reionization in \ion{H}{I} 21 cm gradually diminishes as well. 

As expected, the primary gains in SNR occur at wavenumbers associated with the reionization-bubble scales, and they are minor at small scales. Note that the maximum importance overlaps with the best window in terms of SNR, as shown in Figure \ref{fig:SN_just}. Hence, the cross-correlation of \lya and 21 cm intensity mapping holds significant promise for shedding light on the astrophysics of reionization and it could be used to indirectly constrain the timeline of reionization through its impact in the post-reionization intergalactic medium. Future work will investigate what could be gained from such efforts (see \S\ref{sec:fish} for a preliminary forecast). 

Furthermore, Figure \ref{fig:SN_reio} demonstrates that the impact of inhomogeneous reionization on the post-reionization IGM will increase the SNR for all timelines of reionization and redshifts considered here. In fact, the enhancement can reach $30\%$ at its peak. However, we underscore that there is a fundamental competition in $P_{21, F}$ between the response of underdense regions (forest) and that of denser regions (intensity mapping) to the reionization process. It is plausible that the overall effect would be a reduction of the signal-to-noise at redshifts close to the tail end of reionization. Fortunately, further scrutiny of this hypothesis will not require high-mass resolution simulations like the ones used throughout this work since the role of the small-scale structure should be subdued at these redshifts. Nevertheless, the feasibility of such a measurement is at best uncertain given the current capabilities of spectroscopic instruments and the sparse density of quasars at such high redshifts.  

We summarize our findings regarding observation strategy and dependence with reionization history in terms of signal-to-noise ratio in Table \ref{tab:snr}.

\begin{table}
    \centering
    \caption{Summary of the total SNR at $z = 3.61$ for the different scenarios considered in this work. The Fiducial scenario corresponds to 100 hours, 100 square degrees of survey area, and a Planck-like reionization scenario. We highlight that increasing the survey area by tenfold results in a marginal increase to the SNR for the baseline instrumental setup (see \S\ref{ssec:snr-survey}).}
    \begin{tabular}{c|c|c|c}
        \hline\hline
        \multicolumn{4}{|c|}{\textbf{Total SNR}} \\ 
        {} & SKA $\times$ DESI & SKA $\times$ MUST & PUMA $\times$ MUST \\
        \hline
        Fid. & 8.50 & 13.9 & 110 \\
        \hline
        \multicolumn{4}{|c|}{Increase integration time}\\
        \hline
        1000 h & 21.9 & 35.9 & 167 \\
        \hline
        \multicolumn{4}{|c|}{Increase survey area}\\ 
        \hline
        1000 deg$^2$ & 8.89 & 14.6 & 187 \\
        \hline
        \multicolumn{4}{|c|}{Reionization}\\
        \hline
        None & 7.76 & 12.8 & 103 \\
        Early & 8.10 & 13.3 & 106 \\
        Late & 8.75 & 14.3 & 115 \\
        \hline \hline
    \end{tabular}
    \label{tab:snr}
\end{table}

\section{A simple forecast}
\label{sec:fish}
Having explored the potential gains in SNR due to different strategies and instrumental setups, we now turn our attention to quantifying the potential cosmological gains using the Fisher Matrix formalism \citep{2009arXiv0906.0664H}.

Since no other work has looked at the cross-correlation of \lya $\times$ 21 cm at this redshift range, we will forecast based on three different scenarios: \textit{(i)} Conventional signal, i.e. no memory of reionization, \textit{(ii)} memory of reionization with perfect reionization knowledge, and \textit{(iii)} marginalizing over the memory of reionization due to uncertainty regarding its timeline/modeling. Hence, we can illustrate the potential advantage gained by the reionization relics but simultaneously demonstrate the dangers of ignoring this effect. Furthermore, note that our strategies do not include any sort of priors from other datasets\footnote{One could for instance follow \S2.4 of \cite{2023MNRAS.525.6036L} to incorporate CMB priors; however, this would not address the degeneracy between $\sigma_8$/$n_s$ and $\zeta$. Instead, one could consider using the recent results from \cite{2024arXiv240513680M} to simultaneously break that degeneracy.}, and hence future forecasts are likely to improve in these projections. 

The Fisher matrix is given by
\begin{eqnarray}
    \label{eq:fish}
    F_{\alpha \beta} = \displaystyle \sum_{i}^{z-{\rm bins}}\sum_{j}^{k-{\rm bins}} \sigma^{-2}_{z_ik_j}[P_{21, F}] \frac{\partial P_{21, F}}{\partial \theta_\alpha} (z_i,k_j)  \frac{\partial P_{21, F}}{\partial \theta_\beta}(z_i,k_j) \, \textup{,}
\end{eqnarray}
where we have assumed uncorrelated errors between the different bins and that the posterior distribution can be reasonably well-described by a Gaussian. For simplicity, we consider only two cosmological parameters to forecast  $\theta_i = \{\sigma_8, n_s\}$ in \textit{(i)} and \textit{(ii)}. For \textit{(iii)}, the conservative point of view on the progress of the field, we add the ionization efficiency $\zeta$ to $\boldsymbol{\theta}$.

The uncertainty in Eq.~(\ref{eq:fish}) is obtained from Eq.~(\ref{eq:si-x}). Besides, We use the same three redshift bins centered at 3.61, 3.77, and 3.94 used throughout the rest of this paper. For the wavenumbers, we consider twenty logarithmic bins from 0.01 to 1 Mpc$^{-1}$. 

To compute the derivatives in Eq.~(\ref{eq:fish}), we run additional simulations that cover deviations of 3\% in $\sigma_8$ and $n_s$ around their fiducial values of 0.8159 and 0.9667, respectively. Meanwhile, for $\zeta$ we also consider 3\% variation around $\zeta = 30$ but we check the convergence of our choice for the ionization efficiency in Appendix \ref{app:zeta}. Furthermore, we illustrate some of the tendencies for different properties of this simulation suite in Appendix \ref{app:fish-models}.

\begin{figure}
    \centering
    \includegraphics[width=\linewidth]{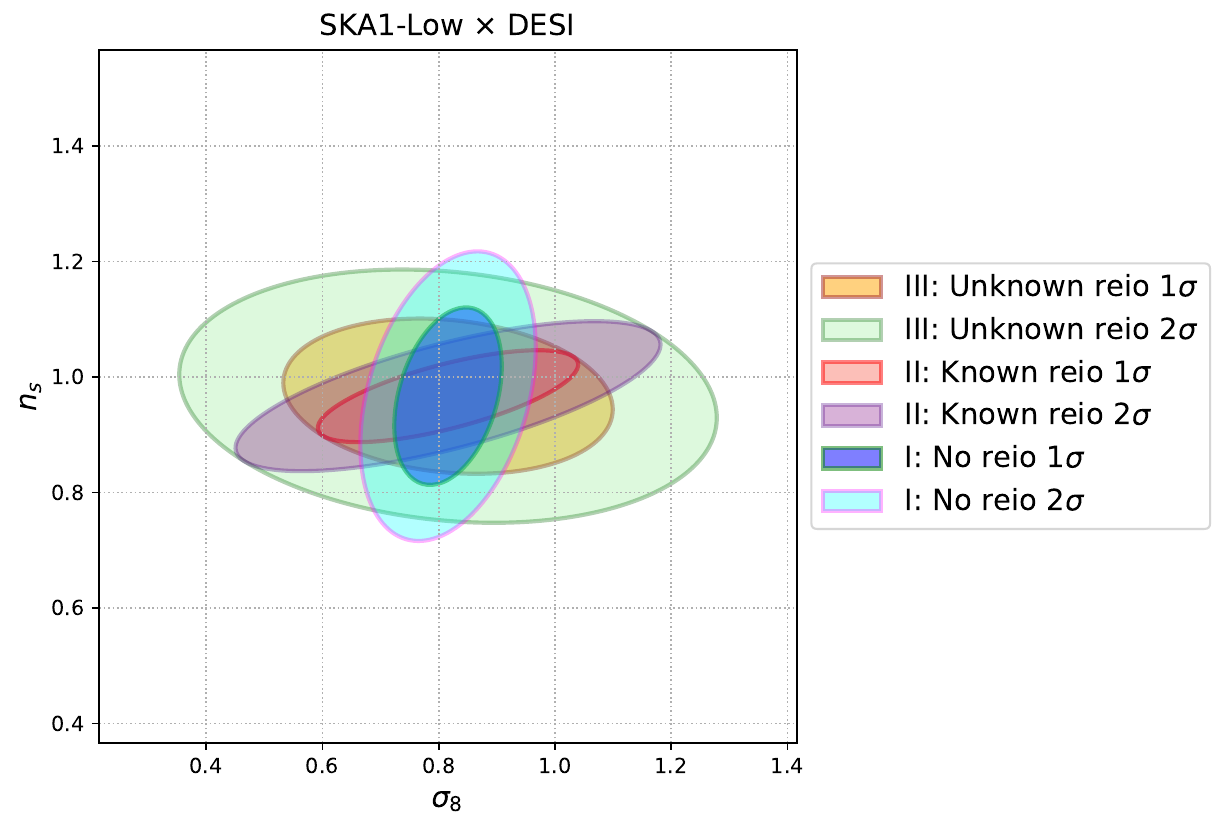}
    \caption{Forecast for the ability of SKA1-Low $\times$ DESI-like to constrain $\sigma_8$ and $n_s$ using 100 sq. degrees and 100 hours of integration time. Shown are the confidence ellipses -- $1\sigma$ and $2\sigma$ -- for different strategies where we consider the case of no reionization (I), i.e. the conventional signal; impact of reionization with known reionization (II), and impact of reionization with uncertain reionization (III). Notably this configuration, which uses only three redshift bins, is not very powerful compared to other cosmological programs expected to be operational by the end of the 2020s; however, it still showcases how reionization relics will jeopardize our ability to constrain cosmology from this observable.}
    \label{fig:fish-ska}
\end{figure}

\begin{table}
    \caption{Projected errors for the Fisher forecast of \S\ref{sec:fish}. Shown are the 1$\sigma$ errors obtained for two instrumental configurations (SKA1-Low $\times$ DESI-like and PUMA $\times$ MUST-like) and for three different scenarios that correspond to no reionization (I), perfect knowledge of reionization/no marginalization over EoR astrophysics (II), and marginalization over ionization efficiency (III), respectively. Both instrumental setups have 100 hours of integration time and 100 sq. deg. of sky coverage.}
    \centering
    \begin{tabular}{c|c|c|c|c|c|c|}
    \hline\hline
        \multicolumn{7}{|c|}{\textbf{Forecasted errors}}\\
        {} & \multicolumn{3}{|c|}{SKA $\times$ DESI} & \multicolumn{3}{|c|}{PUMA $\times$ MUST}\\
        \hline
        Strategy & $\sigma_{\sigma_8}$ & $\sigma_{n_s}$ & $\sigma_\zeta$ & $\sigma_{\sigma_8}$ & $\sigma_{n_s}$ & $\sigma_\zeta$  \\
        \hline
        I & 0.0608 & 0.1011 & --- & 0.0065 & 0.0087 & --- \\
        II & 0.1474 & 0.0524 & --- & 0.0065 & 0.0051 & --- \\
        III & 0.1864 & 0.0884 & 30.39 & 0.0189 & 0.0116 & 3.42
 \\
        \hline\hline
    \end{tabular}
    \label{tab:errors}
\end{table}

\begin{figure}
    \centering
    \includegraphics[width=\linewidth]{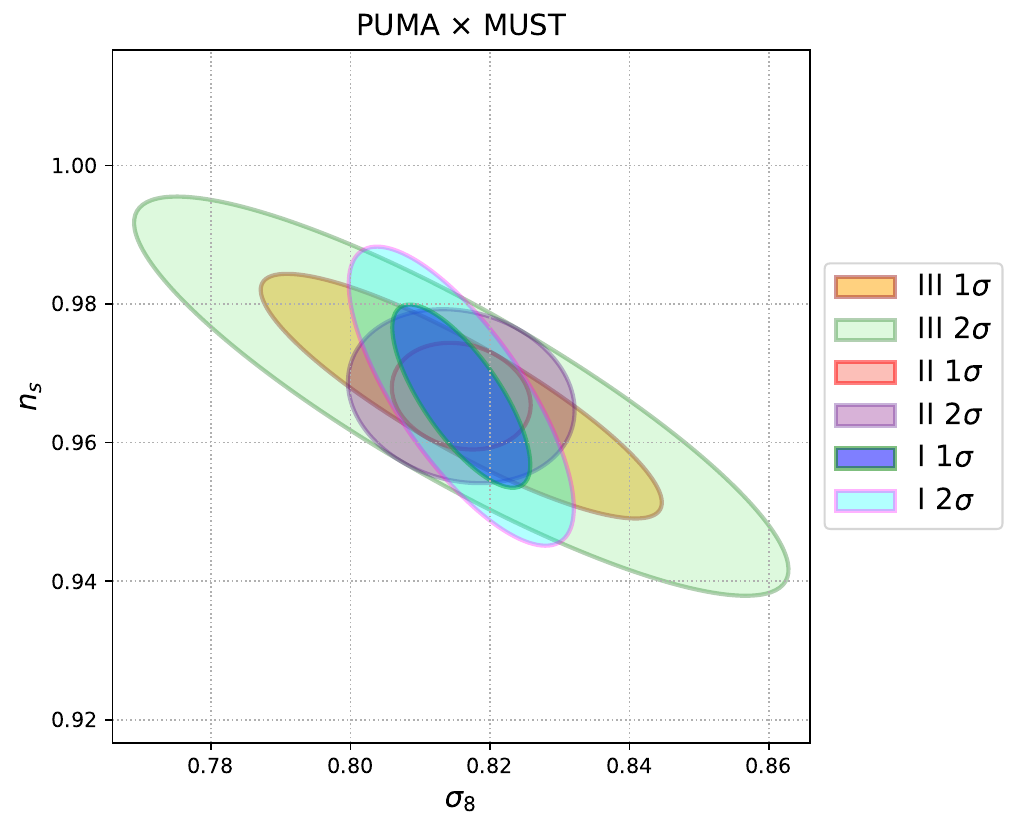}
    \caption{Similar to Figure \ref{fig:fish-ska}, but for PUMA $\times$ MUST-like. In contrast to Figure \ref{fig:fish-ska}, this setup places competitive constraints on the cosmological parameters. Therefore, our results emphasize the importance of considering the impact of reionization to prevent miscalculations of errors and the potential introduction of biases. Note that the combination with other datasets and the use of more than three redshift bins will significantly enhance the real constraining power expected from these instruments.}
    \label{fig:fish-pu}
\end{figure}

We plot the results of the SKA $\times$ DESI-like (PUMA $\times$ MUST-like) forecast in Figure \ref{fig:fish-ska} (Figure \ref{fig:fish-pu}) for the three different strategies considered in this work. The confidence ellipses correspond to the 1 and 2$\sigma$ contours for the $n_s$ $\times$ $\sigma_8$ plane. In addition, we tabulate the projected errors in Table \ref{tab:errors}. 

Both instrumental configurations exhibit a trend of greater ability to constrain the tilt of the primordial power spectrum when the impact of reionization is accounted for and known but when marginalization over the ionization efficiency is required the constraint power for the tilt degrades for the PUMA $\times$ MUST-like configuration. This trend of better $n_s$ constraints was also present in the forecast for \lya forest done by \cite{2021MNRAS.508.1262M}. As seen in their Figures 9 and 10, which include the memory of reionization in the 3D flux power spectrum from the \lya forest, the tilt plays a more important role than the amplitude. We attribute this to the increase (or decrease) of faint galaxies that would happen by modifying the value of $n_s$ while a similar increase in $\sigma_8$ will affect the environment more uniformly, hence influencing the ionizing sources to a lesser degree than the tilt. Note that reionization is driven by faint galaxies in our {\sc 21cmFAST} simulations. The larger significance of the impact of the tilt is also shown in Figure \ref{fig:fish-models} where the impact of changing $n_s$ and $\sigma_8$ in the reionization history is illustrated. 

In the absence of reionization relics, the cross-correlation -- without any external information from other cosmological probes -- seems to constrain $\sigma_8$ more than $n_s$. Interestingly, this is the opposite behavior of the results for \ion{H}{I} 21 cm intensity mapping auto power spectrum albeit with $A_s$ instead (see the 1$\sigma$ error in Table 4 of \citealt{2023MNRAS.525.6036L}). Disregarding the difference of $\sigma_8$ with $A_s$, this trend could be due to the use of Planck priors in the auto-correlation, the additional parameters (like $\Omega$'s that will likely be degenerate with $A_s$ and $\sigma_8$), or perhaps due to the anti-correlated nature of the cross-correlation since in the low-density regions the forest is sensitive to, an increase in $\sigma_8$ would result in enhanced structure formation, and consequently a lesser \lya flux due to the increased absorption. This extra absorption or lack of could play a significant role in the range where the SNR is not dominated by the 21 cm error,i.e. where the \lya forest dominates the error budget, as seen on large scales for the PUMA $\times$ MUST configuration. In these regions, the enhanced sensitivity to the \lya forest increases the constraining power on $\sigma_8$, which could explain the trend of stronger constraint on $\sigma_8$ for the cross-correlation relative to the trend in the autocorrelation.

The \lya $\times$ 21 cm cross-correlation will eventually become a competitive probe of the astrophysics that governs cosmic reionization. However, the constraining power for 100 hours of integration time is too weak to offer real insights into the reionization process. This research direction is likely to become promising with a more ambitious cross-correlation program, say 1000 hours in a SKA $\times$ MUST-like setup. Unexpectedly, even a conservative 100-hour cross-correlation survey will be highly competitive once PUMA \citep{2018arXiv181009572C} starts taking data. For reference, 100 hours of integration time would greatly improve on the projected error for a full DESI (5 years) constraint using the 3D flux power spectrum ($\sigma_\zeta = 11.6$, \citealt{2021MNRAS.508.1262M}) by roughly a factor of 3
. Its constraining power would be of similar strength to that obtained by demoting the optical depth to reionization to derived-parameter using symbolic regression and CMB data in conjunction with astrophysical data \citep{2024arXiv240513680M}.  

\subsection{Constraint on the timeline of reionization}
\label{ssec:time}
Here, we use the results of our PUMA $\times$ MUST-like Strategy III forecast -- marginalize over reionization astrophysics -- to constrain the timeline of reionization, assuming our Fiducial model correctly represents the Universe's neutral hydrogen fraction. In essence, we assume $\zeta = 30$ and we construct the range of allowed $\overline{x}_{\ion{H}{I}}$ based on our forecasted error $\sigma_\zeta$. 

For clarity, we decided to only consider the PUMA configuration, which is significantly stronger than the SKA1-Low $\times$ DESI-like constraint (the error is smaller by almost a factor of 10). This stronger constraining power also justifies our choice of only considering $\zeta$ due to the very small errors for $\sigma_8$ and $n_s$ expected from this instrumental configuration. Additionally, note that our choice of $\zeta = 30$ physically aligns with a scenario where a significant population of faint galaxies is responsible for driving the reionization process -- see the \textit{faint galaxies} model introduced in \cite{2018IAUS..333...18G}.

\begin{figure}
    \centering
    \includegraphics[width=\linewidth]{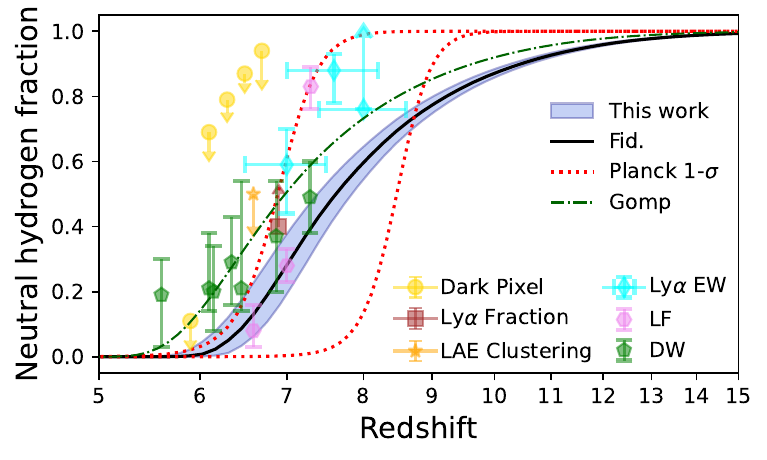}
    \caption{Inferred constraint on the reionization history by our PUMA $\times$ MUST setup using 100 hours of integration and assuming a 100 sq. deg. of overlap. Also shown are current observational constraints on the timeline of reionization including dark pixel fraction \citep{2015MNRAS.447..499M,2023ApJ...942...59J}, high-redshift galaxies through their clustering, luminosity evolution (LF) and equivalent width (EW) \citep{2010ApJ...723..869O,2015MNRAS.453.1843S,2015MNRAS.446..566M,2018ApJ...856....2M, 2019ApJ...878...12H, 2019MNRAS.485.3947M,2021ApJ...919..120M}, and high-$z$ quasars damping wings (DW) \citep{2022MNRAS.512.5390G,2024MNRAS.530.3208G,2024A&A...688L..26S,2024ApJ...969..162D}. We also include the Planck indirect constraints on the timeline of reionization \citep{2020A&A...641A...6P}, which relies on a hyperbolic tangent to parameterize the reionization process. In addition, we include the best fit using Gompertz reionization \citep{2024arXiv240513680M}, which uses Planck data but does not rely on the hyperbolic tangent parametrization and provides a better fit to astrophysical observations. Note that we have anchored the constraint around the fiducial model, thus the key feature is the width of the constraint, rather than its exact location.}
    \label{fig:time}
\end{figure}

We plot our findings in Figure \ref{fig:time} along with the current state of direct and indirect observations (i.e. optical depth constraint from the cosmic microwave background). For our fiducial model, with an ionization efficiency of $\zeta=30$, the ionization of neutral hydrogen begins gradually around $z\sim12$ as denser regions start forming sources of ultraviolet photons capable of ionizing hydrogen. As time progresses, more \ion{H}{II} bubbles form, expand, and merge, leading to an asymmetric reionization process in which the later stages proceed more rapidly. This contrasts with the symmetric sigmoid model, based on a hyperbolic tangent function, used in e.g. \cite{2020A&A...641A...6P} constraint to model reionization.

We emphasize that the width of our inferred constraint is the true value added by our forecast. Our fiducial value, based on the default in {\sc 21cmFASTv3} and aligned with Planck's inferred reionization history\citep{2020A&A...641A...6P}, provides a reasonable description of the Universe's reionization timeline. However, recent advancements in the more reliable quasar damping wing observations \citep[see e.g. ][]{2024ApJ...969..162D,2024MNRAS.530.3208G} suggest a slight tension with both our fiducial model and the Planck constraint. Notably, these developments also support the Gompertz reionization scenario \citep{2024arXiv240513680M}, which successfully fits both CMB data and astrophysical constraints on reionization. Future work should consider a fiducial model that better aligns with the Gompertz timeline. 

Our forecast indicates that measuring the \lya $\times$ 21 cm cross-correlation will likely become a powerful new tool to constrain the evolution of the neutral hydrogen fraction. Such a tight constraint as the one in Figure \ref{fig:time} will impact the standard cosmological model ($\Lambda$CDM), not only by refining the optical depth but also through its dependence on the other cosmological parameters that, in turn, influence the reionization process. Moreover, if observational constraints based on the evolution of the luminosity function of high-$z$ galaxies (represented by pink hexagons in Figure \ref{fig:time}) remain unchanged, the narrowness of our $\overline{x}_{\rm HI} (z)$ constraint could suggest a steeper transition at $z>7$, or perhaps shed light on tensions that may arise between luminosity function and quasar damping wing constraints. Furthermore, we emphasize that reionization is sensitive to alternative cosmological models, such as warm dark matter \citep[e.g.,][]{2014MNRAS.438.2664S}, making strong constraints on the reionization timeline a valuable venue to advance our understanding of dark matter -- and any other processes that may impact cosmic reionization.

We note that our forecast is conservative due to the exclusion of cosmological information from CMB data. At the same time, our choice to use a single astrophysical parameter ($\zeta$) to model reionization is optimistic. The reionization timeline depends on the ease with which ultraviolet photons escape from their host galaxies, i.e. the ionization efficiency $\zeta$. However, the timeline, especially its speed, is also influenced by the ease of forming the parent galaxies of the ionizing photons, often parametrized by a minimum mass threshold $M_{\rm min}$. Naturally, including an additional free astrophysical parameter in the forecast would widen the error estimates in Table \ref{tab:errors}, as some of the constraining power would be redirected to account for the extra degree of freedom. As demonstrated in Figure 11 of \cite{2021MNRAS.508.1262M}, the degeneracy between the two astrophysical parameters are significant. Consequently, we expect the error bars in Figure \ref{fig:time} to increase. While this expected increase in the uncertainty of the $\overline{x}_{\rm HI}(z)$ constraint is a concern, we anticipate that the inclusion of CMB data will more than offset this effect.

\section{Summary}
\label{sec:sum}
The cross-correlation between the \lya forest and 21 cm intensity mapping in the post-reionization era is a promising cosmological probe of the relatively high-redshift intergalactic medium. In particular, it can probe smaller scales before they become fully non-linear (compared to that of traditional galaxy surveys) and it is pragmatically an easier measurement than the auto-correlation of any of those fields (at $z \sim 4$) due to foregrounds or available quasars line of sight. However, just as the \lya forest \citep{2019MNRAS.487.1047M} and 21 cm intensity mapping \citep{2023MNRAS.525.6036L} are sensitive to relics from cosmic reionization, their cross-correlation will also be biased unless appropriate care is taken to handle this broadband effect.    

Regardless of the impact of reionization, our results demonstrate the importance of overlap between radio interferometers and Stage V spectroscopic instruments. Since some of these telescopes are currently in the early planning/design stage, we underscore the significance of guaranteeing a small degree of overlap between the instruments. As shown in Figure \ref{fig:SN_sky}, even 100 square degrees of overlap could lead to a detection and consequently, it would enhance observational programs aimed at the post-reionzation era.  

Furthermore, we found that the gain in signal-to-noise is small for increased survey area in instrumental setups that use SKA1-Low. This trend is caused by the increase in the SKA1-Low thermal noise. Nonetheless, the PUMA $\times$ MUST setup does exhibit significant gains, particularly at large scales with increased survey area, since the error is not dominated by thermal noise. In contrast, a longer integration time leads to considerably better measurements across the board, particularly for SKA configurations. Thus, we conclude that our baseline scenario, which we consider our \textit{cheap} option with 100 hours of integration time and 100 square degrees of overlap, is a well-suited observational setup although the measurements can likely be improved by further adjusting the survey strategy. For instance, a SKA1-Low $\times$ Stage V spectroscopic instrument is likely to perform comparatively to a Stage II radio interferometer at large scales with 1000 hours of integration time (see Figure \ref{fig:SN_tint}). We emphasize that this result highlights the importance of progressing further than Stage V spectroscopic instruments in the 2040s.       

The inclusion of reionization relics in the cross-correlation increases the strength of the signal, particularly at high-$z$ and large scales. Interestingly, late reionization produces an enhancement of up to $30\%$ over the predicted level without reionization imprints at $z\approx 4$. Separating this novel effect from the cosmological information -- for instance using physics-inspired templates \citep{2023MNRAS.520.4853M} -- will allow for unbiased inference of cosmological parameters and it would unseal an original methodology to investigate the astrophysics of reionization. Future work will focus on mitigation strategies for this observable.

We have demonstrated the expected impact of reionization in the inference of cosmological parameters using a Fisher forecast. Cosmological parameters would be biased if one neglects the reionization relics and there would also be a significant underestimation of the error. However, if our knowledge of the reionization timeline improves significantly, the inclusion of the memory of reionization can result in a stronger constraint (as was the case for our PUMA $\times$ MUST forecast in Figure \ref{fig:fish-pu}). Furthermore, the cross-correlation of the \lya forest and \ion{H}{I} 21 cm IM will be a promising probe of reionization astrophysics in the next decades. 

Our findings should also be interpreted as a cautionary tale -- but simultaneously an exciting opportunity -- to other post-reionization era high-$z$ ($z \gtrsim 3$) tracers, like \lya emission \citep[e.g. ][]{2016MNRAS.457.3541C,2021MNRAS.501.3883R,2024arXiv240618775R} and the CO rotational transitions \citep{2022ApJ...929...30B,2022ApJ...933..186C}. Future work will assess the impact of reionization relics in other cosmological tracers of the post-reionization era.

\section*{Acknowledgements}
We thank the anonymous referee for their insightful suggestions. We are grateful to Chris Hirata and Yifan Zheng for their helpful suggestions and comments.
This work was supported by The Major Key Project of PCL. We acknowledge the Tsinghua Astrophysics High-Performance Computing platform at Tsinghua University and PCL's Cloud Brain for providing computational and data storage resources that have contributed to the research results reported within this paper. This work made extensive use of the \hyperlink{https://ui.adsabs.harvard.edu}{NASA Astrophysics DataSystem} and the following open-source python libraries/packages: \texttt{matplotlib} \citep{2007CSE.....9...90H}, \texttt{numpy} \citep{2020Natur.585..357H}, and \texttt{scipy} \citep{2020NatMe..17..261V}.

\section*{Data Availability}
The data underlying this article will be shared on reasonable request to the corresponding authors.



\bibliographystyle{mnras}
\bibliography{cross} 

\begin{thebibliography}{}
\makeatletter
\relax
\def\mn@urlcharsother{\let\do\@makeother \do\$\do\&\do\#\do\^\do\_\do\%\do\~}
\def\mn@doi{\begingroup\mn@urlcharsother \@ifnextchar [ {\mn@doi@}
  {\mn@doi@[]}}
\def\mn@doi@[#1]#2{\def\@tempa{#1}\ifx\@tempa\@empty \href
  {http://dx.doi.org/#2} {doi:#2}\else \href {http://dx.doi.org/#2} {#1}\fi
  \endgroup}
\def\mn@eprint#1#2{\mn@eprint@#1:#2::\@nil}
\def\mn@eprint@arXiv#1{\href {http://arxiv.org/abs/#1} {{\tt arXiv:#1}}}
\def\mn@eprint@dblp#1{\href {http://dblp.uni-trier.de/rec/bibtex/#1.xml}
  {dblp:#1}}
\def\mn@eprint@#1:#2:#3:#4\@nil{\def\@tempa {#1}\def\@tempb {#2}\def\@tempc
  {#3}\ifx \@tempc \@empty \let \@tempc \@tempb \let \@tempb \@tempa \fi \ifx
  \@tempb \@empty \def\@tempb {arXiv}\fi \@ifundefined
  {mn@eprint@\@tempb}{\@tempb:\@tempc}{\expandafter \expandafter \csname
  mn@eprint@\@tempb\endcsname \expandafter{\@tempc}}}

\bibitem[\protect\citeauthoryear{{Abdurashidova} et~al.,}{{Abdurashidova}
  et~al.}{2022}]{2022ApJ...925..221A}
{Abdurashidova} Z.,  et~al., 2022, \mn@doi [\apj] {10.3847/1538-4357/ac1c78},
  \href {https://ui.adsabs.harvard.edu/abs/2022ApJ...925..221A} {925, 221}

\bibitem[\protect\citeauthoryear{{Adams} et~al.,}{{Adams}
  et~al.}{2023}]{2023MNRAS.518.4755A}
{Adams} N.~J.,  et~al., 2023, \mn@doi [\mnras] {10.1093/mnras/stac3347}, \href
  {https://ui.adsabs.harvard.edu/abs/2023MNRAS.518.4755A} {518, 4755}

\bibitem[\protect\citeauthoryear{{Ali} \& {Bharadwaj}}{{Ali} \&
  {Bharadwaj}}{2014}]{2014JApA...35..157A}
{Ali} S.~S.,  {Bharadwaj} S.,  2014, \mn@doi [Journal of Astrophysics and
  Astronomy] {10.1007/s12036-014-9301-1}, \href
  {https://ui.adsabs.harvard.edu/abs/2014JApA...35..157A} {35, 157}

\bibitem[\protect\citeauthoryear{{Amiri} et~al.,}{{Amiri}
  et~al.}{2023}]{2023ApJ...947...16A}
{Amiri} M.,  et~al., 2023, \mn@doi [\apj] {10.3847/1538-4357/acb13f}, \href
  {https://ui.adsabs.harvard.edu/abs/2023ApJ...947...16A} {947, 16}

\bibitem[\protect\citeauthoryear{{Arinyo-i-Prats}, {Miralda-Escud{\'e}}, {Viel}
   \& {Cen}}{{Arinyo-i-Prats} et~al.}{2015}]{2015JCAP...12..017A}
{Arinyo-i-Prats} A.,  {Miralda-Escud{\'e}} J.,  {Viel} M.,   {Cen} R.,  2015,
  \mn@doi [\jcap] {10.1088/1475-7516/2015/12/017}, \href
  {https://ui.adsabs.harvard.edu/abs/2015JCAP...12..017A} {2015, 017}

\bibitem[\protect\citeauthoryear{{Atek} et~al.,}{{Atek}
  et~al.}{2023}]{2023MNRAS.519.1201A}
{Atek} H.,  et~al., 2023, \mn@doi [\mnras] {10.1093/mnras/stac3144}, \href
  {https://ui.adsabs.harvard.edu/abs/2023MNRAS.519.1201A} {519, 1201}

\bibitem[\protect\citeauthoryear{{Bharadwaj}, {Sarkar}  \& {Ali}}{{Bharadwaj}
  et~al.}{2015}]{2015JApA...36..385B}
{Bharadwaj} S.,  {Sarkar} A.~K.,   {Ali} S.~S.,  2015, \mn@doi [Journal of
  Astrophysics and Astronomy] {10.1007/s12036-015-9346-9}, \href
  {https://ui.adsabs.harvard.edu/abs/2015JApA...36..385B} {36, 385}

\bibitem[\protect\citeauthoryear{{Bradley} et~al.,}{{Bradley}
  et~al.}{2022}]{2022arXiv221001777B}
{Bradley} L.~D.,  et~al., 2022, \mn@doi [arXiv e-prints]
  {10.48550/arXiv.2210.01777}, \href
  {https://ui.adsabs.harvard.edu/abs/2022arXiv221001777B} {p. arXiv:2210.01777}

\bibitem[\protect\citeauthoryear{{Braun}, {Bonaldi}, {Bourke}, {Keane}  \&
  {Wagg}}{{Braun} et~al.}{2019}]{2019arXiv191212699B}
{Braun} R.,  {Bonaldi} A.,  {Bourke} T.,  {Keane} E.,   {Wagg} J.,  2019,
  \mn@doi [arXiv e-prints] {10.48550/arXiv.1912.12699}, \href
  {https://ui.adsabs.harvard.edu/abs/2019arXiv191212699B} {p. arXiv:1912.12699}

\bibitem[\protect\citeauthoryear{{Breysse}, {Yang}, {Somerville}, {Pullen},
  {Popping}  \& {Maniyar}}{{Breysse} et~al.}{2022}]{2022ApJ...929...30B}
{Breysse} P.~C.,  {Yang} S.,  {Somerville} R.~S.,  {Pullen} A.~R.,  {Popping}
  G.,   {Maniyar} A.~S.,  2022, \mn@doi [\apj] {10.3847/1538-4357/ac5a46},
  \href {https://ui.adsabs.harvard.edu/abs/2022ApJ...929...30B} {929, 30}

\bibitem[\protect\citeauthoryear{{Bull}, {Ferreira}, {Patel}  \&
  {Santos}}{{Bull} et~al.}{2015}]{2015ApJ...803...21B}
{Bull} P.,  {Ferreira} P.~G.,  {Patel} P.,   {Santos} M.~G.,  2015, \mn@doi
  [\apj] {10.1088/0004-637X/803/1/21}, \href
  {https://ui.adsabs.harvard.edu/abs/2015ApJ...803...21B} {803, 21}

\bibitem[\protect\citeauthoryear{{Carucci}, {Villaescusa-Navarro}, {Viel}  \&
  {Lapi}}{{Carucci} et~al.}{2015}]{2015JCAP...07..047C}
{Carucci} I.~P.,  {Villaescusa-Navarro} F.,  {Viel} M.,   {Lapi} A.,  2015,
  \mn@doi [\jcap] {10.1088/1475-7516/2015/07/047}, \href
  {https://ui.adsabs.harvard.edu/abs/2015JCAP...07..047C} {2015, 047}

\bibitem[\protect\citeauthoryear{{Carucci}, {Villaescusa-Navarro}  \&
  {Viel}}{{Carucci} et~al.}{2017}]{2017JCAP...04..001C}
{Carucci} I.~P.,  {Villaescusa-Navarro} F.,   {Viel} M.,  2017, \mn@doi [\jcap]
  {10.1088/1475-7516/2017/04/001}, \href
  {https://ui.adsabs.harvard.edu/abs/2017JCAP...04..001C} {2017, 001}

\bibitem[\protect\citeauthoryear{{Castorina} \&
  {Villaescusa-Navarro}}{{Castorina} \&
  {Villaescusa-Navarro}}{2017}]{2017MNRAS.471.1788C}
{Castorina} E.,  {Villaescusa-Navarro} F.,  2017, \mn@doi [\mnras]
  {10.1093/mnras/stx1599}, \href
  {https://ui.adsabs.harvard.edu/abs/2017MNRAS.471.1788C} {471, 1788}

\bibitem[\protect\citeauthoryear{{Castorina} \& {White}}{{Castorina} \&
  {White}}{2019}]{2019JCAP...06..025C}
{Castorina} E.,  {White} M.,  2019, \mn@doi [\jcap]
  {10.1088/1475-7516/2019/06/025}, \href
  {https://ui.adsabs.harvard.edu/abs/2019JCAP...06..025C} {2019, 025}

\bibitem[\protect\citeauthoryear{{Cen}, {McDonald}, {Trac}  \& {Loeb}}{{Cen}
  et~al.}{2009}]{2009ApJ...706L.164C}
{Cen} R.,  {McDonald} P.,  {Trac} H.,   {Loeb} A.,  2009, \mn@doi [\apjl]
  {10.1088/0004-637X/706/1/L164}, \href
  {http://adsabs.harvard.edu/abs/2009ApJ...706L.164C} {706, L164}

\bibitem[\protect\citeauthoryear{{Chabanier} et~al.,}{{Chabanier}
  et~al.}{2019}]{2019JCAP...07..017C}
{Chabanier} S.,  et~al., 2019, \mn@doi [\jcap] {10.1088/1475-7516/2019/07/017},
  \href {https://ui.adsabs.harvard.edu/abs/2019JCAP...07..017C} {2019, 017}

\bibitem[\protect\citeauthoryear{{Chaussidon} et~al.,}{{Chaussidon}
  et~al.}{2023}]{2023ApJ...944..107C}
{Chaussidon} E.,  et~al., 2023, \mn@doi [\apj] {10.3847/1538-4357/acb3c2},
  \href {https://ui.adsabs.harvard.edu/abs/2023ApJ...944..107C} {944, 107}

\bibitem[\protect\citeauthoryear{{Choudhury}, {Haehnelt}  \&
  {Regan}}{{Choudhury} et~al.}{2009}]{2009MNRAS.394..960C}
{Choudhury} T.~R.,  {Haehnelt} M.~G.,   {Regan} J.,  2009, \mn@doi [\mnras]
  {10.1111/j.1365-2966.2008.14383.x}, \href
  {https://ui.adsabs.harvard.edu/abs/2009MNRAS.394..960C} {394, 960}

\bibitem[\protect\citeauthoryear{{Chung} et~al.,}{{Chung}
  et~al.}{2022}]{2022ApJ...933..186C}
{Chung} D.~T.,  et~al., 2022, \mn@doi [\apj] {10.3847/1538-4357/ac63c7}, \href
  {https://ui.adsabs.harvard.edu/abs/2022ApJ...933..186C} {933, 186}

\bibitem[\protect\citeauthoryear{{Cosmic Visions 21 cm Collaboration}
  et~al.,}{{Cosmic Visions 21 cm Collaboration}
  et~al.}{2018}]{2018arXiv181009572C}
{Cosmic Visions 21 cm Collaboration} et~al., 2018, arXiv e-prints, \href
  {https://ui.adsabs.harvard.edu/abs/2018arXiv181009572C} {p. arXiv:1810.09572}

\bibitem[\protect\citeauthoryear{{Crighton} et~al.,}{{Crighton}
  et~al.}{2015}]{2015MNRAS.452..217C}
{Crighton} N. H.~M.,  et~al., 2015, \mn@doi [\mnras] {10.1093/mnras/stv1182},
  \href {https://ui.adsabs.harvard.edu/abs/2015MNRAS.452..217C} {452, 217}

\bibitem[\protect\citeauthoryear{{Croft} et~al.,}{{Croft}
  et~al.}{2016}]{2016MNRAS.457.3541C}
{Croft} R. A.~C.,  et~al., 2016, \mn@doi [\mnras] {10.1093/mnras/stw204}, \href
  {https://ui.adsabs.harvard.edu/abs/2016MNRAS.457.3541C} {457, 3541}

\bibitem[\protect\citeauthoryear{{DESI Collaboration} et~al.,}{{DESI
  Collaboration} et~al.}{2022}]{2022AJ....164..207D}
{DESI Collaboration} et~al., 2022, \mn@doi [\aj] {10.3847/1538-3881/ac882b},
  \href {https://ui.adsabs.harvard.edu/abs/2022AJ....164..207D} {164, 207}

\bibitem[\protect\citeauthoryear{{DESI Collaboration} et~al.,}{{DESI
  Collaboration} et~al.}{2024}]{2024arXiv240403001D}
{DESI Collaboration} et~al., 2024, \mn@doi [arXiv e-prints]
  {10.48550/arXiv.2404.03001}, \href
  {https://ui.adsabs.harvard.edu/abs/2024arXiv240403001D} {p. arXiv:2404.03001}

\bibitem[\protect\citeauthoryear{{Dash} \& {Guha Sarkar}}{{Dash} \& {Guha
  Sarkar}}{2021}]{2021JCAP...02..016D}
{Dash} C. B.~V.,  {Guha Sarkar} T.,  2021, \mn@doi [\jcap]
  {10.1088/1475-7516/2021/02/016}, \href
  {https://ui.adsabs.harvard.edu/abs/2021JCAP...02..016D} {2021, 016}

\bibitem[\protect\citeauthoryear{{Dash}, {Sarkar}  \& {Sarkar}}{{Dash}
  et~al.}{2023}]{2023JApA...44....5D}
{Dash} C. B.~V.,  {Sarkar} T.~G.,   {Sarkar} A.~K.,  2023, \mn@doi [Journal of
  Astrophysics and Astronomy] {10.1007/s12036-022-09885-w}, \href
  {https://ui.adsabs.harvard.edu/abs/2023JApA...44....5D} {44, 5}

\bibitem[\protect\citeauthoryear{{Dayal} \& {Ferrara}}{{Dayal} \&
  {Ferrara}}{2018}]{2018PhR...780....1D}
{Dayal} P.,  {Ferrara} A.,  2018, \mn@doi [\physrep]
  {10.1016/j.physrep.2018.10.002}, \href
  {https://ui.adsabs.harvard.edu/\#abs/2018PhR...780....1D} {780, 1}

\bibitem[\protect\citeauthoryear{{Diao}, {Grumitt}  \& {Mao}}{{Diao}
  et~al.}{2024}]{2024arXiv240711296D}
{Diao} K.,  {Grumitt} R. D.~P.,   {Mao} Y.,  2024, \mn@doi [arXiv e-prints]
  {10.48550/arXiv.2407.11296}, \href
  {https://ui.adsabs.harvard.edu/abs/2024arXiv240711296D} {p. arXiv:2407.11296}

\bibitem[\protect\citeauthoryear{{Donnan} et~al.,}{{Donnan}
  et~al.}{2023}]{2023MNRAS.518.6011D}
{Donnan} C.~T.,  et~al., 2023, \mn@doi [\mnras] {10.1093/mnras/stac3472}, \href
  {https://ui.adsabs.harvard.edu/abs/2023MNRAS.518.6011D} {518, 6011}

\bibitem[\protect\citeauthoryear{{Facchinetti}, {Lopez-Honorez}, {Qin}  \&
  {Mesinger}}{{Facchinetti} et~al.}{2023}]{2023arXiv230816656F}
{Facchinetti} G.,  {Lopez-Honorez} L.,  {Qin} Y.,   {Mesinger} A.,  2023,
  \mn@doi [arXiv e-prints] {10.48550/arXiv.2308.16656}, \href
  {https://ui.adsabs.harvard.edu/abs/2023arXiv230816656F} {p. arXiv:2308.16656}

\bibitem[\protect\citeauthoryear{{Filbert} et~al.,}{{Filbert}
  et~al.}{2023}]{2023arXiv230903434F}
{Filbert} S.,  et~al., 2023, \mn@doi [arXiv e-prints]
  {10.48550/arXiv.2309.03434}, \href
  {https://ui.adsabs.harvard.edu/abs/2023arXiv230903434F} {p. arXiv:2309.03434}

\bibitem[\protect\citeauthoryear{{Font-Ribera}, {McDonald}, {Mostek}, {Reid},
  {Seo}  \& {Slosar}}{{Font-Ribera} et~al.}{2014}]{2014JCAP...05..023F}
{Font-Ribera} A.,  {McDonald} P.,  {Mostek} N.,  {Reid} B.~A.,  {Seo} H.-J.,
  {Slosar} A.,  2014, \mn@doi [\jcap] {10.1088/1475-7516/2014/05/023}, \href
  {https://ui.adsabs.harvard.edu/abs/2014JCAP...05..023F} {2014, 023}

\bibitem[\protect\citeauthoryear{{Furlanetto} \& {Lidz}}{{Furlanetto} \&
  {Lidz}}{2007}]{2007ApJ...660.1030F}
{Furlanetto} S.~R.,  {Lidz} A.,  2007, \mn@doi [\apj] {10.1086/513009}, \href
  {https://ui.adsabs.harvard.edu/abs/2007ApJ...660.1030F} {660, 1030}

\bibitem[\protect\citeauthoryear{{Gardner} et~al.,}{{Gardner}
  et~al.}{2006}]{2006SSRv..123..485G}
{Gardner} J.~P.,  et~al., 2006, \mn@doi [\ssr] {10.1007/s11214-006-8315-7},
  \href {https://ui.adsabs.harvard.edu/abs/2006SSRv..123..485G} {123, 485}

\bibitem[\protect\citeauthoryear{{Gnedin}}{{Gnedin}}{2022}]{2022ApJ...937...17G}
{Gnedin} N.~Y.,  2022, \mn@doi [\apj] {10.3847/1538-4357/ac8d0a}, \href
  {https://ui.adsabs.harvard.edu/abs/2022ApJ...937...17G} {937, 17}

\bibitem[\protect\citeauthoryear{{Gordon} et~al.,}{{Gordon}
  et~al.}{2023}]{2023JCAP...11..045G}
{Gordon} C.,  et~al., 2023, \mn@doi [\jcap] {10.1088/1475-7516/2023/11/045},
  \href {https://ui.adsabs.harvard.edu/abs/2023JCAP...11..045G} {2023, 045}

\bibitem[\protect\citeauthoryear{{Greig} \& {Mesinger}}{{Greig} \&
  {Mesinger}}{2018}]{2018IAUS..333...18G}
{Greig} B.,  {Mesinger} A.,  2018, in {Jeli{\'c}} V.,  {van der Hulst} T.,
  eds,  IAU Symposium Vol. 333, Peering towards Cosmic Dawn. pp 18--21
  (\mn@eprint {arXiv} {1705.03471}), \mn@doi{10.1017/S1743921317011103}

\bibitem[\protect\citeauthoryear{{Greig}, {Mesinger}, {Davies}, {Wang}, {Yang}
  \& {Hennawi}}{{Greig} et~al.}{2022}]{2022MNRAS.512.5390G}
{Greig} B.,  {Mesinger} A.,  {Davies} F.~B.,  {Wang} F.,  {Yang} J.,
  {Hennawi} J.~F.,  2022, \mn@doi [\mnras] {10.1093/mnras/stac825}, \href
  {https://ui.adsabs.harvard.edu/abs/2022MNRAS.512.5390G} {512, 5390}

\bibitem[\protect\citeauthoryear{{Greig} et~al.,}{{Greig}
  et~al.}{2024}]{2024MNRAS.530.3208G}
{Greig} B.,  et~al., 2024, \mn@doi [\mnras] {10.1093/mnras/stae1080}, \href
  {https://ui.adsabs.harvard.edu/abs/2024MNRAS.530.3208G} {530, 3208}

\bibitem[\protect\citeauthoryear{{Guha Sarkar} \& {Datta}}{{Guha Sarkar} \&
  {Datta}}{2015}]{2015JCAP...08..001G}
{Guha Sarkar} T.,  {Datta} K.~K.,  2015, \mn@doi [\jcap]
  {10.1088/1475-7516/2015/08/001}, \href
  {https://ui.adsabs.harvard.edu/abs/2015JCAP...08..001G} {2015, 001}

\bibitem[\protect\citeauthoryear{{Guha Sarkar}, {Bharadwaj}, {Choudhury}  \&
  {Datta}}{{Guha Sarkar} et~al.}{2011}]{2011MNRAS.410.1130G}
{Guha Sarkar} T.,  {Bharadwaj} S.,  {Choudhury} T.~R.,   {Datta} K.~K.,  2011,
  \mn@doi [\mnras] {10.1111/j.1365-2966.2010.17509.x}, \href
  {https://ui.adsabs.harvard.edu/abs/2011MNRAS.410.1130G} {410, 1130}

\bibitem[\protect\citeauthoryear{{Guy} et~al.,}{{Guy}
  et~al.}{2023}]{2023AJ....165..144G}
{Guy} J.,  et~al., 2023, \mn@doi [\aj] {10.3847/1538-3881/acb212}, \href
  {https://ui.adsabs.harvard.edu/abs/2023AJ....165..144G} {165, 144}

\bibitem[\protect\citeauthoryear{{Harris} et~al.,}{{Harris}
  et~al.}{2020}]{2020Natur.585..357H}
{Harris} C.~R.,  et~al., 2020, \mn@doi [\nat] {10.1038/s41586-020-2649-2},
  \href {https://ui.adsabs.harvard.edu/abs/2020Natur.585..357H} {585, 357}

\bibitem[\protect\citeauthoryear{{Heavens}}{{Heavens}}{2009}]{2009arXiv0906.0664H}
{Heavens} A.,  2009, \mn@doi [arXiv e-prints] {10.48550/arXiv.0906.0664}, \href
  {https://ui.adsabs.harvard.edu/abs/2009arXiv0906.0664H} {p. arXiv:0906.0664}

\bibitem[\protect\citeauthoryear{{Hirata}}{{Hirata}}{2018}]{2018MNRAS.474.2173H}
{Hirata} C.~M.,  2018, \mn@doi [\mnras] {10.1093/mnras/stx2854}, \href
  {https://ui.adsabs.harvard.edu/abs/2018MNRAS.474.2173H} {474, 2173}

\bibitem[\protect\citeauthoryear{{Hoag} et~al.,}{{Hoag}
  et~al.}{2019}]{2019ApJ...878...12H}
{Hoag} A.,  et~al., 2019, \mn@doi [\apj] {10.3847/1538-4357/ab1de7}, \href
  {https://ui.adsabs.harvard.edu/abs/2019ApJ...878...12H} {878, 12}

\bibitem[\protect\citeauthoryear{{Hunter}}{{Hunter}}{2007}]{2007CSE.....9...90H}
{Hunter} J.~D.,  2007, \mn@doi [Computing in Science and Engineering]
  {10.1109/MCSE.2007.55}, \href
  {https://ui.adsabs.harvard.edu/abs/2007CSE.....9...90H} {9, 90}

\bibitem[\protect\citeauthoryear{{Hutter}, {Heneka}, {Dayal}, {Gottl{\"o}ber},
  {Mesinger}, {Trebitsch}  \& {Yepes}}{{Hutter}
  et~al.}{2023}]{2023MNRAS.525.1664H}
{Hutter} A.,  {Heneka} C.,  {Dayal} P.,  {Gottl{\"o}ber} S.,  {Mesinger} A.,
  {Trebitsch} M.,   {Yepes} G.,  2023, \mn@doi [\mnras]
  {10.1093/mnras/stad2376}, \href
  {https://ui.adsabs.harvard.edu/abs/2023MNRAS.525.1664H} {525, 1664}

\bibitem[\protect\citeauthoryear{{Jin} et~al.,}{{Jin}
  et~al.}{2023}]{2023ApJ...942...59J}
{Jin} X.,  et~al., 2023, \mn@doi [\apj] {10.3847/1538-4357/aca678}, \href
  {https://ui.adsabs.harvard.edu/abs/2023ApJ...942...59J} {942, 59}

\bibitem[\protect\citeauthoryear{{Karim}, {Armengaud}, {Mention}, {Chabanier},
  {Ravoux}  \& {Luki{\'c}}}{{Karim} et~al.}{2023}]{2023arXiv231009116K}
{Karim} M. L.~A.,  {Armengaud} E.,  {Mention} G.,  {Chabanier} S.,  {Ravoux}
  C.,   {Luki{\'c}} Z.,  2023, \mn@doi [arXiv e-prints]
  {10.48550/arXiv.2310.09116}, \href
  {https://ui.adsabs.harvard.edu/abs/2023arXiv231009116K} {p. arXiv:2310.09116}

\bibitem[\protect\citeauthoryear{{Keating}, {Weinberger}, {Kulkarni},
  {Haehnelt}, {Chardin}  \& {Aubert}}{{Keating}
  et~al.}{2020}]{2020MNRAS.491.1736K}
{Keating} L.~C.,  {Weinberger} L.~H.,  {Kulkarni} G.,  {Haehnelt} M.~G.,
  {Chardin} J.,   {Aubert} D.,  2020, \mn@doi [\mnras] {10.1093/mnras/stz3083},
  \href {https://ui.adsabs.harvard.edu/abs/2020MNRAS.491.1736K} {491, 1736}

\bibitem[\protect\citeauthoryear{{Kim}, {Bolton}, {Viel}, {Haehnelt}  \&
  {Carswell}}{{Kim} et~al.}{2007}]{2007MNRAS.382.1657K}
{Kim} T.~S.,  {Bolton} J.~S.,  {Viel} M.,  {Haehnelt} M.~G.,   {Carswell}
  R.~F.,  2007, \mn@doi [\mnras] {10.1111/j.1365-2966.2007.12406.x}, \href
  {https://ui.adsabs.harvard.edu/abs/2007MNRAS.382.1657K} {382, 1657}

\bibitem[\protect\citeauthoryear{{La Plante}, {Mirocha}, {Gorce}, {Lidz}  \&
  {Parsons}}{{La Plante} et~al.}{2023}]{2023ApJ...944...59L}
{La Plante} P.,  {Mirocha} J.,  {Gorce} A.,  {Lidz} A.,   {Parsons} A.,  2023,
  \mn@doi [\apj] {10.3847/1538-4357/acaeb0}, \href
  {https://ui.adsabs.harvard.edu/abs/2023ApJ...944...59L} {944, 59}

\bibitem[\protect\citeauthoryear{{Lee}, {Cen}, {Gott}  \& {Trac}}{{Lee}
  et~al.}{2008}]{2008ApJ...675....8L}
{Lee} K.-G.,  {Cen} R.,  {Gott} J.~Richard I.,   {Trac} H.,  2008, \mn@doi
  [\apj] {10.1086/525520}, \href
  {https://ui.adsabs.harvard.edu/abs/2008ApJ...675....8L} {675, 8}

\bibitem[\protect\citeauthoryear{{Lewis}}{{Lewis}}{2008}]{2008PhRvD..78b3002L}
{Lewis} A.,  2008, \mn@doi [\prd] {10.1103/PhysRevD.78.023002}, \href
  {https://ui.adsabs.harvard.edu/abs/2008PhRvD..78b3002L} {78, 023002}

\bibitem[\protect\citeauthoryear{{Long} \& {Hirata}}{{Long} \&
  {Hirata}}{2023}]{2023MNRAS.520..948L}
{Long} H.,  {Hirata} C.~M.,  2023, \mn@doi [\mnras] {10.1093/mnras/stad184},
  \href {https://ui.adsabs.harvard.edu/abs/2023MNRAS.520..948L} {520, 948}

\bibitem[\protect\citeauthoryear{{Long}, {Givans}  \& {Hirata}}{{Long}
  et~al.}{2022}]{2022MNRAS.513..117L}
{Long} H.,  {Givans} J.~J.,   {Hirata} C.~M.,  2022, \mn@doi [\mnras]
  {10.1093/mnras/stac658}, \href
  {https://ui.adsabs.harvard.edu/abs/2022MNRAS.513..117L} {513, 117}

\bibitem[\protect\citeauthoryear{{Long}, {Morales-Guti{\'e}rrez},
  {Montero-Camacho}  \& {Hirata}}{{Long} et~al.}{2023}]{2023MNRAS.525.6036L}
{Long} H.,  {Morales-Guti{\'e}rrez} C.,  {Montero-Camacho} P.,   {Hirata}
  C.~M.,  2023, \mn@doi [\mnras] {10.1093/mnras/stad2639}, \href
  {https://ui.adsabs.harvard.edu/abs/2023MNRAS.525.6036L} {525, 6036}

\bibitem[\protect\citeauthoryear{{Mason}, {Treu}, {Dijkstra}, {Mesinger},
  {Trenti}, {Pentericci}, {de Barros}  \& {Vanzella}}{{Mason}
  et~al.}{2018}]{2018ApJ...856....2M}
{Mason} C.~A.,  {Treu} T.,  {Dijkstra} M.,  {Mesinger} A.,  {Trenti} M.,
  {Pentericci} L.,  {de Barros} S.,   {Vanzella} E.,  2018, \mn@doi [\apj]
  {10.3847/1538-4357/aab0a7}, \href
  {https://ui.adsabs.harvard.edu/abs/2018ApJ...856....2M} {856, 2}

\bibitem[\protect\citeauthoryear{{Mason} et~al.,}{{Mason}
  et~al.}{2019}]{2019MNRAS.485.3947M}
{Mason} C.~A.,  et~al., 2019, \mn@doi [\mnras] {10.1093/mnras/stz632}, \href
  {https://ui.adsabs.harvard.edu/abs/2019MNRAS.485.3947M} {485, 3947}

\bibitem[\protect\citeauthoryear{{McDonald} \& {Eisenstein}}{{McDonald} \&
  {Eisenstein}}{2007}]{2007PhRvD..76f3009M}
{McDonald} P.,  {Eisenstein} D.~J.,  2007, \mn@doi [\prd]
  {10.1103/PhysRevD.76.063009}, \href
  {https://ui.adsabs.harvard.edu/abs/2007PhRvD..76f3009M} {76, 063009}

\bibitem[\protect\citeauthoryear{{McGreer}, {Mesinger}  \&
  {D'Odorico}}{{McGreer} et~al.}{2015}]{2015MNRAS.447..499M}
{McGreer} I.~D.,  {Mesinger} A.,   {D'Odorico} V.,  2015, \mn@doi [\mnras]
  {10.1093/mnras/stu2449}, \href
  {https://ui.adsabs.harvard.edu/abs/2015MNRAS.447..499M} {447, 499}

\bibitem[\protect\citeauthoryear{{Mertens} et~al.,}{{Mertens}
  et~al.}{2020}]{2020MNRAS.493.1662M}
{Mertens} F.~G.,  et~al., 2020, \mn@doi [\mnras] {10.1093/mnras/staa327}, \href
  {https://ui.adsabs.harvard.edu/abs/2020MNRAS.493.1662M} {493, 1662}

\bibitem[\protect\citeauthoryear{{Mesinger}}{{Mesinger}}{2016}]{2016ASSL..423.....M}
{Mesinger} A.,  2016, {Understanding the Epoch of Cosmic Reionization}.
 Astrophysics and Space Science Library Vol. 423,
  \mn@doi{10.1007/978-3-319-21957-8, }

\bibitem[\protect\citeauthoryear{{Mesinger}}{{Mesinger}}{2019}]{2019cosm.book.....M}
{Mesinger} A.,  2019, {The Cosmic 21-cm Revolution; Charting the first billion
  years of our universe}, \mn@doi{10.1088/2514-3433/ab4a73.
}

\bibitem[\protect\citeauthoryear{{Mesinger}, {Furlanetto}  \& {Cen}}{{Mesinger}
  et~al.}{2011}]{2011MNRAS.411..955M}
{Mesinger} A.,  {Furlanetto} S.,   {Cen} R.,  2011, \mn@doi [\mnras]
  {10.1111/j.1365-2966.2010.17731.x}, \href
  {https://ui.adsabs.harvard.edu/abs/2011MNRAS.411..955M} {411, 955}

\bibitem[\protect\citeauthoryear{{Mesinger}, {Aykutalp}, {Vanzella},
  {Pentericci}, {Ferrara}  \& {Dijkstra}}{{Mesinger}
  et~al.}{2015}]{2015MNRAS.446..566M}
{Mesinger} A.,  {Aykutalp} A.,  {Vanzella} E.,  {Pentericci} L.,  {Ferrara} A.,
    {Dijkstra} M.,  2015, \mn@doi [\mnras] {10.1093/mnras/stu2089}, \href
  {https://ui.adsabs.harvard.edu/abs/2015MNRAS.446..566M} {446, 566}

\bibitem[\protect\citeauthoryear{{Montero-Camacho} \& {Mao}}{{Montero-Camacho}
  \& {Mao}}{2020}]{2020MNRAS.499.1640M}
{Montero-Camacho} P.,  {Mao} Y.,  2020, \mn@doi [\mnras]
  {10.1093/mnras/staa2918}, \href
  {https://ui.adsabs.harvard.edu/abs/2020MNRAS.499.1640M} {499, 1640}

\bibitem[\protect\citeauthoryear{{Montero-Camacho} \& {Mao}}{{Montero-Camacho}
  \& {Mao}}{2021}]{2021MNRAS.508.1262M}
{Montero-Camacho} P.,  {Mao} Y.,  2021, \mn@doi [\mnras]
  {10.1093/mnras/stab2569}, \href
  {https://ui.adsabs.harvard.edu/abs/2021MNRAS.508.1262M} {508, 1262}

\bibitem[\protect\citeauthoryear{{Montero-Camacho}, {Hirata}, {Martini}  \&
  {Honscheid}}{{Montero-Camacho} et~al.}{2019}]{2019MNRAS.487.1047M}
{Montero-Camacho} P.,  {Hirata} C.~M.,  {Martini} P.,   {Honscheid} K.,  2019,
  \mn@doi [\mnras] {10.1093/mnras/stz1388}, \href
  {https://ui.adsabs.harvard.edu/abs/2019MNRAS.487.1047M} {487, 1047}

\bibitem[\protect\citeauthoryear{{Montero-Camacho}, {Liu}  \&
  {Mao}}{{Montero-Camacho} et~al.}{2023}]{2023MNRAS.520.4853M}
{Montero-Camacho} P.,  {Liu} Y.,   {Mao} Y.,  2023, \mn@doi [\mnras]
  {10.1093/mnras/stad437}, \href
  {https://ui.adsabs.harvard.edu/abs/2023MNRAS.520.4853M} {520, 4853}

\bibitem[\protect\citeauthoryear{{Montero-Camacho}, {Li}  \&
  {Cranmer}}{{Montero-Camacho} et~al.}{2024a}]{2024arXiv240513680M}
{Montero-Camacho} P.,  {Li} Y.,   {Cranmer} M.,  2024a, \mn@doi [arXiv
  e-prints] {10.48550/arXiv.2405.13680}, \href
  {https://ui.adsabs.harvard.edu/abs/2024arXiv240513680M} {p. arXiv:2405.13680}

\bibitem[\protect\citeauthoryear{{Montero-Camacho}, {Zhang}  \&
  {Mao}}{{Montero-Camacho} et~al.}{2024b}]{2024MNRAS.529.3666M}
{Montero-Camacho} P.,  {Zhang} Y.,   {Mao} Y.,  2024b, \mn@doi [\mnras]
  {10.1093/mnras/stae751}, \href
  {https://ui.adsabs.harvard.edu/abs/2024MNRAS.529.3666M} {529, 3666}

\bibitem[\protect\citeauthoryear{{Morales}, {Mason}, {Bruton}, {Gronke},
  {Haardt}  \& {Scarlata}}{{Morales} et~al.}{2021}]{2021ApJ...919..120M}
{Morales} A.~M.,  {Mason} C.~A.,  {Bruton} S.,  {Gronke} M.,  {Haardt} F.,
  {Scarlata} C.,  2021, \mn@doi [\apj] {10.3847/1538-4357/ac1104}, \href
  {https://ui.adsabs.harvard.edu/abs/2021ApJ...919..120M} {919, 120}

\bibitem[\protect\citeauthoryear{{Munshi} et~al.,}{{Munshi}
  et~al.}{2023}]{2023arXiv231105364M}
{Munshi} S.,  et~al., 2023, \mn@doi [arXiv e-prints]
  {10.48550/arXiv.2311.05364}, \href
  {https://ui.adsabs.harvard.edu/abs/2023arXiv231105364M} {p. arXiv:2311.05364}

\bibitem[\protect\citeauthoryear{{Murray}, {Greig}, {Mesinger}, {Mu{\~n}oz},
  {Qin}, {Park}  \& {Watkinson}}{{Murray} et~al.}{2020}]{2020JOSS....5.2582M}
{Murray} S.,  {Greig} B.,  {Mesinger} A.,  {Mu{\~n}oz} J.,  {Qin} Y.,  {Park}
  J.,   {Watkinson} C.,  2020, \mn@doi [The Journal of Open Source Software]
  {10.21105/joss.02582}, \href
  {https://ui.adsabs.harvard.edu/abs/2020JOSS....5.2582M} {5, 2582}

\bibitem[\protect\citeauthoryear{{Ouchi} et~al.,}{{Ouchi}
  et~al.}{2010}]{2010ApJ...723..869O}
{Ouchi} M.,  et~al., 2010, \mn@doi [\apj] {10.1088/0004-637X/723/1/869}, \href
  {https://ui.adsabs.harvard.edu/abs/2010ApJ...723..869O} {723, 869}

\bibitem[\protect\citeauthoryear{{Palanque-Delabrouille}
  et~al.,}{{Palanque-Delabrouille} et~al.}{2013a}]{2013A&A...551A..29P}
{Palanque-Delabrouille} N.,  et~al., 2013a, \mn@doi [\aap]
  {10.1051/0004-6361/201220379}, \href
  {https://ui.adsabs.harvard.edu/abs/2013A&A...551A..29P} {551, A29}

\bibitem[\protect\citeauthoryear{{Palanque-Delabrouille}
  et~al.,}{{Palanque-Delabrouille} et~al.}{2013b}]{2013A&A...559A..85P}
{Palanque-Delabrouille} N.,  et~al., 2013b, \mn@doi [\aap]
  {10.1051/0004-6361/201322130}, \href
  {https://ui.adsabs.harvard.edu/abs/2013A&A...559A..85P} {559, A85}

\bibitem[\protect\citeauthoryear{{Palanque-Delabrouille}, {Y{\`e}che},
  {Sch{\"o}neberg}, {Lesgourgues}, {Walther}, {Chabanier}  \&
  {Armengaud}}{{Palanque-Delabrouille} et~al.}{2020}]{2020JCAP...04..038P}
{Palanque-Delabrouille} N.,  {Y{\`e}che} C.,  {Sch{\"o}neberg} N.,
  {Lesgourgues} J.,  {Walther} M.,  {Chabanier} S.,   {Armengaud} E.,  2020,
  \mn@doi [\jcap] {10.1088/1475-7516/2020/04/038}, \href
  {https://ui.adsabs.harvard.edu/abs/2020JCAP...04..038P} {2020, 038}

\bibitem[\protect\citeauthoryear{{Park}, {Mesinger}, {Greig}  \&
  {Gillet}}{{Park} et~al.}{2019}]{2019MNRAS.484..933P}
{Park} J.,  {Mesinger} A.,  {Greig} B.,   {Gillet} N.,  2019, \mn@doi [\mnras]
  {10.1093/mnras/stz032}, \href
  {https://ui.adsabs.harvard.edu/abs/2019MNRAS.484..933P} {484, 933}

\bibitem[\protect\citeauthoryear{{Park}, {Luki{\'c}}, {Sexton}, {Alvarez}  \&
  {Shapiro}}{{Park} et~al.}{2024}]{2024ApJ...969...46P}
{Park} H.,  {Luki{\'c}} Z.,  {Sexton} J.,  {Alvarez} M.~A.,   {Shapiro} P.~R.,
  2024, \mn@doi [\apj] {10.3847/1538-4357/ad4bdc}, \href
  {https://ui.adsabs.harvard.edu/abs/2024ApJ...969...46P} {969, 46}

\bibitem[\protect\citeauthoryear{{Percival} et~al.,}{{Percival}
  et~al.}{2019}]{2019arXiv190303158P}
{Percival} W.~J.,  et~al., 2019, \mn@doi [arXiv e-prints]
  {10.48550/arXiv.1903.03158}, \href
  {https://ui.adsabs.harvard.edu/abs/2019arXiv190303158P} {p. arXiv:1903.03158}

\bibitem[\protect\citeauthoryear{{Planck Collaboration} et~al.,}{{Planck
  Collaboration} et~al.}{2016}]{2016A&A...594A..13P}
{Planck Collaboration} et~al., 2016, \mn@doi [\aap]
  {10.1051/0004-6361/201525830}, \href
  {https://ui.adsabs.harvard.edu/abs/2016A&A...594A..13P} {594, A13}

\bibitem[\protect\citeauthoryear{{Planck Collaboration} et~al.,}{{Planck
  Collaboration} et~al.}{2020}]{2020A&A...641A...6P}
{Planck Collaboration} et~al., 2020, \mn@doi [\aap]
  {10.1051/0004-6361/201833910}, \href
  {https://ui.adsabs.harvard.edu/abs/2020A&A...641A...6P} {641, A6}

\bibitem[\protect\citeauthoryear{{Pober} et~al.,}{{Pober}
  et~al.}{2014}]{2014ApJ...782...66P}
{Pober} J.~C.,  et~al., 2014, \mn@doi [\apj] {10.1088/0004-637X/782/2/66},
  \href {https://ui.adsabs.harvard.edu/abs/2014ApJ...782...66P} {782, 66}

\bibitem[\protect\citeauthoryear{{Puchwein} et~al.,}{{Puchwein}
  et~al.}{2022}]{2022MNRAS.tmp.3519P}
{Puchwein} E.,  et~al., 2022, \mn@doi [\mnras] {10.1093/mnras/stac3761}, \href
  {https://ui.adsabs.harvard.edu/abs/2022MNRAS.tmp.3519P} {}

\bibitem[\protect\citeauthoryear{{Ram{\'\i}rez-P{\'e}rez}
  et~al.,}{{Ram{\'\i}rez-P{\'e}rez} et~al.}{2023}]{2023arXiv230606312R}
{Ram{\'\i}rez-P{\'e}rez} C.,  et~al., 2023, \mn@doi [arXiv e-prints]
  {10.48550/arXiv.2306.06312}, \href
  {https://ui.adsabs.harvard.edu/abs/2023arXiv230606312R} {p. arXiv:2306.06312}

\bibitem[\protect\citeauthoryear{{Ravoux} et~al.,}{{Ravoux}
  et~al.}{2023}]{2023MNRAS.526.5118R}
{Ravoux} C.,  et~al., 2023, \mn@doi [\mnras] {10.1093/mnras/stad3008}, \href
  {https://ui.adsabs.harvard.edu/abs/2023MNRAS.526.5118R} {526, 5118}

\bibitem[\protect\citeauthoryear{{Renard} et~al.,}{{Renard}
  et~al.}{2021}]{2021MNRAS.501.3883R}
{Renard} P.,  et~al., 2021, \mn@doi [\mnras] {10.1093/mnras/staa3783}, \href
  {https://ui.adsabs.harvard.edu/abs/2021MNRAS.501.3883R} {501, 3883}

\bibitem[\protect\citeauthoryear{{Renard}, {Spinoso}, {Sun}, {Zou},
  {Montero-Camacho}  \& {Cai}}{{Renard} et~al.}{2024}]{2024arXiv240618775R}
{Renard} P.,  {Spinoso} D.,  {Sun} Z.,  {Zou} H.,  {Montero-Camacho} P.,
  {Cai} Z.,  2024, \mn@doi [arXiv e-prints] {10.48550/arXiv.2406.18775}, \href
  {https://ui.adsabs.harvard.edu/abs/2024arXiv240618775R} {p. arXiv:2406.18775}

\bibitem[\protect\citeauthoryear{{Richard} et~al.,}{{Richard}
  et~al.}{2019}]{2019Msngr.175...50R}
{Richard} J.,  et~al., 2019, \mn@doi [The Messenger] {10.18727/0722-6691/5127},
  \href {https://ui.adsabs.harvard.edu/abs/2019Msngr.175...50R} {175, 50}

\bibitem[\protect\citeauthoryear{{Roth}, {D'Aloisio}, {Cain}, {Wilson}, {Zhu}
  \& {Becker}}{{Roth} et~al.}{2023}]{2023arXiv231106348R}
{Roth} J.~T.,  {D'Aloisio} A.,  {Cain} C.,  {Wilson} B.,  {Zhu} Y.,   {Becker}
  G.~D.,  2023, \mn@doi [arXiv e-prints] {10.48550/arXiv.2311.06348}, \href
  {https://ui.adsabs.harvard.edu/abs/2023arXiv231106348R} {p. arXiv:2311.06348}

\bibitem[\protect\citeauthoryear{{Sarkar}, {Bharadwaj}  \& {Guha
  Sarkar}}{{Sarkar} et~al.}{2018}]{2018JCAP...05..051S}
{Sarkar} A.~K.,  {Bharadwaj} S.,   {Guha Sarkar} T.,  2018, \mn@doi [\jcap]
  {10.1088/1475-7516/2018/05/051}, \href
  {https://ui.adsabs.harvard.edu/abs/2018JCAP...05..051S} {2018, 051}

\bibitem[\protect\citeauthoryear{{Sarkar}, {Pal}  \& {Guha Sarkar}}{{Sarkar}
  et~al.}{2019}]{2019JCAP...12..058S}
{Sarkar} A.~K.,  {Pal} A.~K.,   {Guha Sarkar} T.,  2019, \mn@doi [\jcap]
  {10.1088/1475-7516/2019/12/058}, \href
  {https://ui.adsabs.harvard.edu/abs/2019JCAP...12..058S} {2019, 058}

\bibitem[\protect\citeauthoryear{{Schlegel}, {Kollmeier}  \&
  {Ferraro}}{{Schlegel} et~al.}{2019}]{2019BAAS...51g.229S}
{Schlegel} D.,  {Kollmeier} J.~A.,   {Ferraro} S.,  2019, in Bulletin of the
  American Astronomical Society. p.~229 (\mn@eprint {arXiv} {1907.11171}),
  \mn@doi{10.48550/arXiv.1907.11171}

\bibitem[\protect\citeauthoryear{{Sitwell}, {Mesinger}, {Ma}  \&
  {Sigurdson}}{{Sitwell} et~al.}{2014}]{2014MNRAS.438.2664S}
{Sitwell} M.,  {Mesinger} A.,  {Ma} Y.-Z.,   {Sigurdson} K.,  2014, \mn@doi
  [\mnras] {10.1093/mnras/stt2392}, \href
  {https://ui.adsabs.harvard.edu/abs/2014MNRAS.438.2664S} {438, 2664}

\bibitem[\protect\citeauthoryear{{Slosar} et~al.,}{{Slosar}
  et~al.}{2019}]{2019BAAS...51g..53S}
{Slosar} A.,  et~al., 2019, in Bulletin of the American Astronomical Society.
  p.~53 (\mn@eprint {arXiv} {1907.12559})

\bibitem[\protect\citeauthoryear{{Sobacchi} \& {Mesinger}}{{Sobacchi} \&
  {Mesinger}}{2015}]{2015MNRAS.453.1843S}
{Sobacchi} E.,  {Mesinger} A.,  2015, \mn@doi [\mnras] {10.1093/mnras/stv1751},
  \href {https://ui.adsabs.harvard.edu/abs/2015MNRAS.453.1843S} {453, 1843}

\bibitem[\protect\citeauthoryear{{Spina}, {Bosman}, {Davies}, {Gaikwad}  \&
  {Zhu}}{{Spina} et~al.}{2024}]{2024A&A...688L..26S}
{Spina} B.,  {Bosman} S. E.~I.,  {Davies} F.~B.,  {Gaikwad} P.,   {Zhu} Y.,
  2024, \mn@doi [\aap] {10.1051/0004-6361/202450798}, \href
  {https://ui.adsabs.harvard.edu/abs/2024A&A...688L..26S} {688, L26}

\bibitem[\protect\citeauthoryear{{Springel}}{{Springel}}{2005}]{2005MNRAS.364.1105S}
{Springel} V.,  2005, \mn@doi [\mnras] {10.1111/j.1365-2966.2005.09655.x},
  \href {http://adsabs.harvard.edu/abs/2005MNRAS.364.1105S} {364, 1105}

\bibitem[\protect\citeauthoryear{{Square Kilometre Array Cosmology Science
  Working Group} et~al.,}{{Square Kilometre Array Cosmology Science Working
  Group} et~al.}{2020}]{2020PASA...37....7S}
{Square Kilometre Array Cosmology Science Working Group} et~al., 2020, \mn@doi
  [\pasa] {10.1017/pasa.2019.51}, \href
  {https://ui.adsabs.harvard.edu/abs/2020PASA...37....7S} {37, e007}

\bibitem[\protect\citeauthoryear{{Sun}, {Ting}  \& {Cai}}{{Sun}
  et~al.}{2023}]{2023ApJS..269....4S}
{Sun} Z.,  {Ting} Y.-S.,   {Cai} Z.,  2023, \mn@doi [\apjs]
  {10.3847/1538-4365/acf2f1}, \href
  {https://ui.adsabs.harvard.edu/abs/2023ApJS..269....4S} {269, 4}

\bibitem[\protect\citeauthoryear{{Upton Sanderbeck} \& {Bird}}{{Upton
  Sanderbeck} \& {Bird}}{2020}]{2020MNRAS.496.4372U}
{Upton Sanderbeck} P.,  {Bird} S.,  2020, \mn@doi [\mnras]
  {10.1093/mnras/staa1850}, \href
  {https://ui.adsabs.harvard.edu/abs/2020MNRAS.496.4372U} {496, 4372}

\bibitem[\protect\citeauthoryear{{Villaescusa-Navarro}, {Viel}, {Alonso},
  {Datta}, {Bull}  \& {Santos}}{{Villaescusa-Navarro}
  et~al.}{2015}]{2015JCAP...03..034V}
{Villaescusa-Navarro} F.,  {Viel} M.,  {Alonso} D.,  {Datta} K.~K.,  {Bull} P.,
    {Santos} M.~G.,  2015, \mn@doi [\jcap] {10.1088/1475-7516/2015/03/034},
  \href {https://ui.adsabs.harvard.edu/abs/2015JCAP...03..034V} {2015, 034}

\bibitem[\protect\citeauthoryear{{Virtanen} et~al.,}{{Virtanen}
  et~al.}{2020}]{2020NatMe..17..261V}
{Virtanen} P.,  et~al., 2020, \mn@doi [Nature Methods]
  {10.1038/s41592-019-0686-2}, \href
  {https://ui.adsabs.harvard.edu/abs/2020NatMe..17..261V} {17, 261}

\bibitem[\protect\citeauthoryear{{Visbal}, {Loeb}  \& {Wyithe}}{{Visbal}
  et~al.}{2009}]{2009JCAP...10..030V}
{Visbal} E.,  {Loeb} A.,   {Wyithe} S.,  2009, \mn@doi [\jcap]
  {10.1088/1475-7516/2009/10/030}, \href
  {https://ui.adsabs.harvard.edu/abs/2009JCAP...10..030V} {2009, 030}

\bibitem[\protect\citeauthoryear{{Wang} et~al.,}{{Wang}
  et~al.}{2022}]{2022ApJS..259...28W}
{Wang} B.,  et~al., 2022, \mn@doi [\apjs] {10.3847/1538-4365/ac4504}, \href
  {https://ui.adsabs.harvard.edu/abs/2022ApJS..259...28W} {259, 28}

\bibitem[\protect\citeauthoryear{{Weinberg}, {Dav{\'e}}, {Katz}  \&
  {Kollmeier}}{{Weinberg} et~al.}{2003}]{2003AIPC..666..157W}
{Weinberg} D.~H.,  {Dav{\'e}} R.,  {Katz} N.,   {Kollmeier} J.~A.,  2003, in
  {Holt} S.~H.,  {Reynolds} C.~S.,  eds,  American Institute of Physics
  Conference Series Vol. 666, The Emergence of Cosmic Structure. pp 157--169
  (\mn@eprint {} {astro-ph/0301186}), \mn@doi{10.1063/1.1581786}

\bibitem[\protect\citeauthoryear{{Wolz} et~al.,}{{Wolz}
  et~al.}{2015}]{2015aska.confE..35W}
{Wolz} L.,  et~al., 2015, in Advancing Astrophysics with the Square Kilometre
  Array (AASKA14). p.~35 (\mn@eprint {arXiv} {1501.03823}),
  \mn@doi{10.22323/1.215.0035}

\bibitem[\protect\citeauthoryear{{Wyithe} \& {Loeb}}{{Wyithe} \&
  {Loeb}}{2009}]{2009MNRAS.397.1926W}
{Wyithe} J. S.~B.,  {Loeb} A.,  2009, \mn@doi [\mnras]
  {10.1111/j.1365-2966.2009.15019.x}, \href
  {https://ui.adsabs.harvard.edu/abs/2009MNRAS.397.1926W} {397, 1926}

\bibitem[\protect\citeauthoryear{{Yang} et~al.,}{{Yang}
  et~al.}{2023}]{2023arXiv230201777Y}
{Yang} J.,  et~al., 2023, \mn@doi [arXiv e-prints] {10.48550/arXiv.2302.01777},
  \href {https://ui.adsabs.harvard.edu/abs/2023arXiv230201777Y} {p.
  arXiv:2302.01777}

\bibitem[\protect\citeauthoryear{{Y{\`e}che}, {Palanque-Delabrouille}, {Baur}
  \& {du Mas des Bourboux}}{{Y{\`e}che} et~al.}{2017}]{2017JCAP...06..047Y}
{Y{\`e}che} C.,  {Palanque-Delabrouille} N.,  {Baur} J.,   {du Mas des
  Bourboux} H.,  2017, \mn@doi [\jcap] {10.1088/1475-7516/2017/06/047}, \href
  {http://adsabs.harvard.edu/abs/2017JCAP...06..047Y} {6, 047}

\bibitem[\protect\citeauthoryear{{Y{\`e}che} et~al.,}{{Y{\`e}che}
  et~al.}{2020}]{2020RNAAS...4..179Y}
{Y{\`e}che} C.,  et~al., 2020, \mn@doi [Research Notes of the American
  Astronomical Society] {10.3847/2515-5172/abc01a}, \href
  {https://ui.adsabs.harvard.edu/abs/2020RNAAS...4..179Y} {4, 179}

\bibitem[\protect\citeauthoryear{{Zhang} et~al.,}{{Zhang}
  et~al.}{2024}]{2024MNRAS.530.1235Z}
{Zhang} Y.,  et~al., 2024, \mn@doi [\mnras] {10.1093/mnras/stae871}, \href
  {https://ui.adsabs.harvard.edu/abs/2024MNRAS.530.1235Z} {530, 1235}

\bibitem[\protect\citeauthoryear{{Zhao}, {Mao}  \& {Wandelt}}{{Zhao}
  et~al.}{2022}]{2022ApJ...933..236Z}
{Zhao} X.,  {Mao} Y.,   {Wandelt} B.~D.,  2022, \mn@doi [\apj]
  {10.3847/1538-4357/ac778e}, \href
  {https://ui.adsabs.harvard.edu/abs/2022ApJ...933..236Z} {933, 236}

\bibitem[\protect\citeauthoryear{{Zhou}, {Tan}  \& {Mao}}{{Zhou}
  et~al.}{2021}]{2021ApJ...909...51Z}
{Zhou} M.,  {Tan} J.,   {Mao} Y.,  2021, \mn@doi [\apj]
  {10.3847/1538-4357/abda45}, \href
  {https://ui.adsabs.harvard.edu/abs/2021ApJ...909...51Z} {909, 51}

\bibitem[\protect\citeauthoryear{{Zuo}, {Chen}  \& {Mao}}{{Zuo}
  et~al.}{2023}]{2023ApJ...945...38Z}
{Zuo} S.,  {Chen} X.,   {Mao} Y.,  2023, \mn@doi [\apj]
  {10.3847/1538-4357/acb822}, \href
  {https://ui.adsabs.harvard.edu/abs/2023ApJ...945...38Z} {945, 38}

\bibitem[\protect\citeauthoryear{{du Mas des Bourboux} et~al.,}{{du Mas des
  Bourboux} et~al.}{2020}]{2020ApJ...901..153D}
{du Mas des Bourboux} H.,  et~al., 2020, \mn@doi [\apj]
  {10.3847/1538-4357/abb085}, \href
  {https://ui.adsabs.harvard.edu/abs/2020ApJ...901..153D} {901, 153}

\bibitem[\protect\citeauthoryear{{{\v{D}}urov{\v{c}}{\'\i}kov{\'a}}
  et~al.,}{{{\v{D}}urov{\v{c}}{\'\i}kov{\'a}}
  et~al.}{2024}]{2024ApJ...969..162D}
{{\v{D}}urov{\v{c}}{\'\i}kov{\'a}} D.,  et~al., 2024, \mn@doi [\apj]
  {10.3847/1538-4357/ad4888}, \href
  {https://ui.adsabs.harvard.edu/abs/2024ApJ...969..162D} {969, 162}

\makeatother
\end{thebibliography}



\appendix
\section{Dependence of forecast on $\delta \zeta$}
\label{app:zeta}
The ionization efficiency $\zeta$ controls the timing of reionization in {\sc 21cmFAST} by parametrizing the difficulty for an ionizing photon to escape into the IGM. Consequently, this parameter is not as well constrained or studied as cosmological parameters like $\sigma_8$ or $n_s$. Here we confirm that the step size of this uncertain parameter does not have a significant impact on our results. We follow \cite{2023arXiv230816656F} strategy of checking the impact on the forecasted error instead of the performance of the numerical derivative. 

\begin{figure*}
    \centering
    \includegraphics[width=\linewidth]{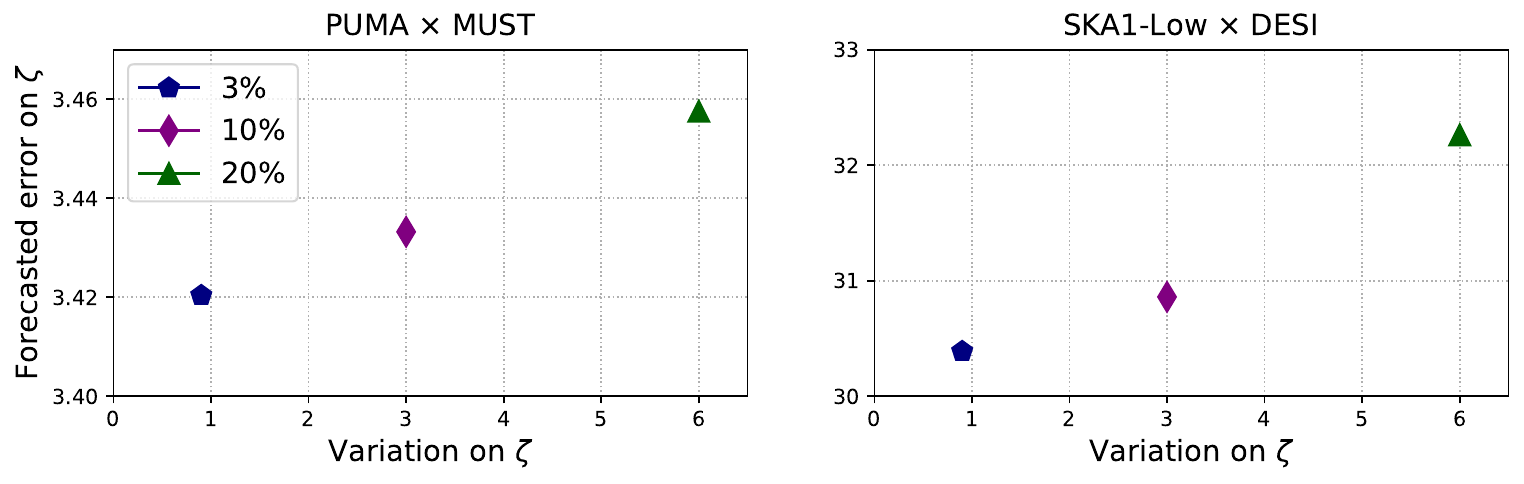}
    \caption{Convergence of the ionization efficiency error with step size. Shown are the forecasted errors for a variation of 3\% (blue pentagons), 10\% (purple diamonds), and 20\% (green triangles) of the fiducial $\zeta$. Given the small difference between the forecasted errors, we opt for the 3\% variation in ionization efficiency.}
    \label{fig:zeta}
\end{figure*}

Figure \ref{fig:zeta} highlights the small difference between the forecasted $\zeta$ errors and justifies our choice of 3\% variation in ionization efficiency for our Fisher forecast. 

\section{Trends for the Fisher simulations}
\label{app:fish-models}
Here we show some reionization-related properties of the simulations used in \S\ref{sec:fish} to clarify their hierarchy. 

\begin{figure*}
    \centering
    \includegraphics[width=\linewidth]{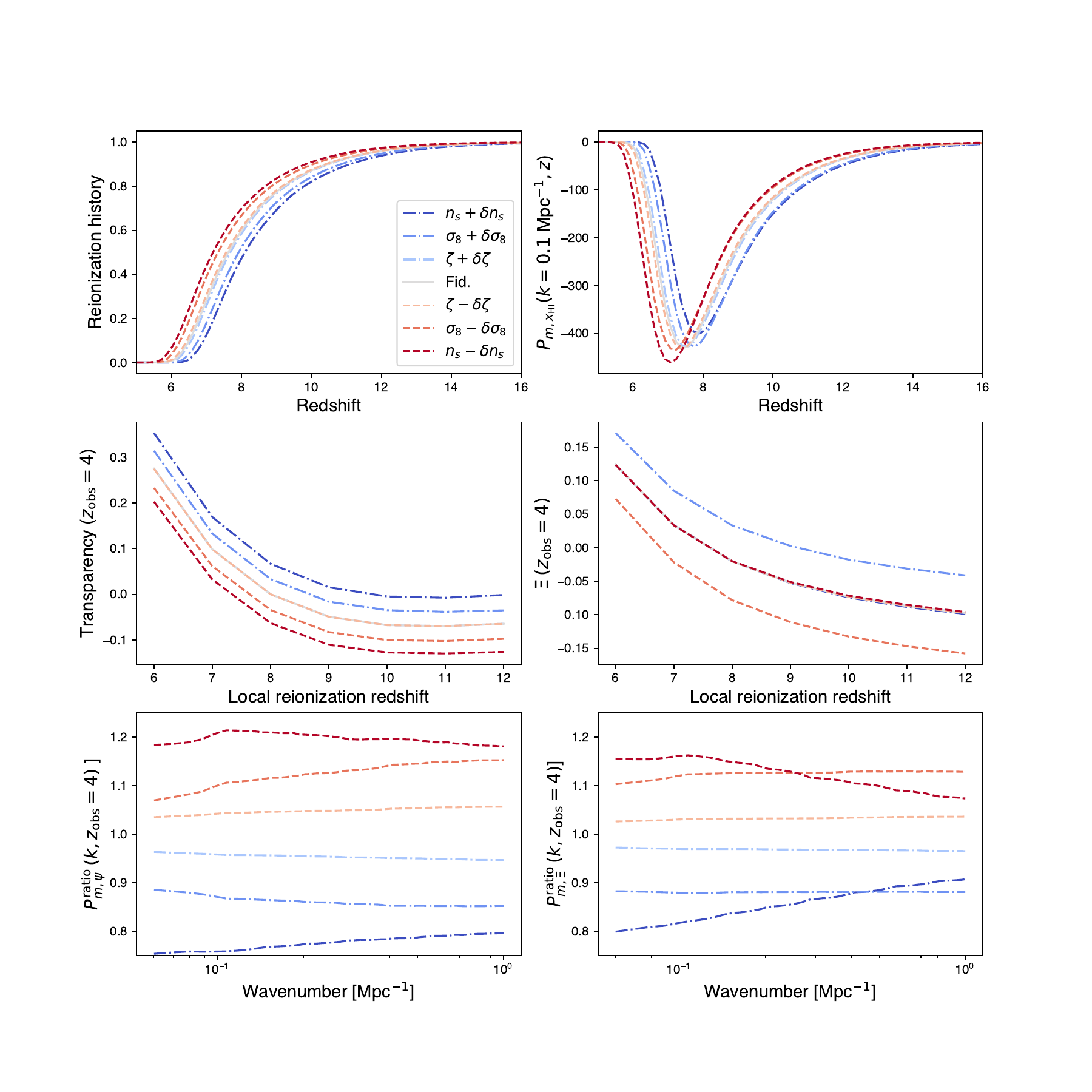}
    \caption{Properties of the simulations used in the Fisher forecast of section \ref{sec:fish}. (Top left) Reionization history. (Top right) Cross-correlation of matter and neutral hydrogen field as a function of redshift and evaluated at a $k$ representative of the ionized bubble scale. (Middle left) Relative transparency of the IGM $\psi$ as a function of local reionization redshift, Eq.~(\ref{eq:psi}). (Middle right) Relative neutral hydrogen density $\Xi$, Eq.~(\ref{eq:xi}) (Bottom left) The cross-power spectrum of matter and transparency of the model divided by that of the fiducial simulation, i.e. Eq.~(\ref{eq:psi_ref}) divided by the reference scenario. (Bottom right) Ratio of the cross-power spectrum of matter and relative neutral hydrogen density, i.e. Eq.~(\ref{eq:xi_ref}) normalized by the reference scenario.}
    \label{fig:fish-models}
\end{figure*}

For instance, it is clear that $n_s$ will be the most constrained parameter by the inclusion of reionization relics since it dominates most of the different metrics in Figure \ref{fig:fish-models}. In contrast, a weaker constraint in ionization efficiency is also expected, particularly since ionization efficiency does not affect the first term in Eq.~(\ref{eq:power}).

\bsp	
\label{lastpage}
\end{document}